\pdfoutput=1

\documentclass[aps,prl,superscriptaddress,amsmath,amssymb,twocolumn]{revtex4-2}

\usepackage{graphicx,bm,amsmath}
\usepackage[usenames,dvipsnames]{xcolor}
\usepackage[colorlinks,bookmarks=false,citecolor=NavyBlue,linkcolor=Red,urlcolor=blue,
]{hyperref}
\usepackage{physics}
\usepackage[percent]{overpic}
\usepackage[bbgreekl]{mathbbol}
\usepackage{xcolor}
\usepackage[normalem]{ulem}
\usepackage{algorithm}
\usepackage{braket}
\usepackage{algpseudocode}
\usepackage{comment}
\usepackage{amsthm}
\usepackage{enumerate}

\usepackage{algorithm}
\usepackage{algpseudocode}

\def\doi{http://dx.doi.org/}

\newcommand{\be}{\begin{equation}}
\newcommand{\ee}{\end{equation}}
\newcommand{\bec}{\begin{equation*}}
\newcommand{\eec}{\end{equation*}}
\newcommand{\bea}{\begin{eqnarray}}
\newcommand{\eea}{\end{eqnarray}}

\usepackage{graphicx}
\usepackage{dcolumn}
\usepackage{bm}

\begin{document}

\title{Nonstabilizerness in U(1) lattice gauge theory} 

\author{Pedro R. Nic\'acio Falc\~ao}
\email{pedro.nicaciofalcao@doctoral.uj.edu.pl}
\affiliation{Szkoła Doktorska Nauk \'Scis\l{}ych i Przyrodniczych, Uniwersytet Jagiello\'nski,  \L{}ojasiewicza 11, PL-30-348 Krak\'ow, Poland}
\affiliation{Instytut Fizyki Teoretycznej, Wydzia\l{} Fizyki, Astronomii i Informatyki Stosowanej,
Uniwersytet Jagiello\'nski,  \L{}ojasiewicza 11, PL-30-348 Krak\'ow, Poland}

\author{Poetri Sonya Tarabunga}
\email{ptarabun@sissa.it}
\affiliation{The Abdus Salam International Centre for Theoretical Physics (ICTP), Strada Costiera 11, 34151 Trieste, Italy}
\affiliation{International School for Advanced Studies (SISSA), Via Bonomea 265, I-34136 Trieste, Italy}
\affiliation{INFN, Sezione di Trieste, Via Valerio 2, 34127 Trieste, Italy}

\author{Martina Frau}
\affiliation{International School for Advanced Studies (SISSA), Via Bonomea 265, I-34136 Trieste, Italy}

\author{Emanuele Tirrito}
\email{etirrito@ictp.it}
\affiliation{The Abdus Salam International Centre for Theoretical Physics (ICTP), Strada Costiera 11, 34151 Trieste, Italy}
\affiliation{Pitaevskii BEC Center, CNR-INO and Dipartimento di Fisica,
Università di Trento, Via Sommarive 14, Trento, I-38123, Italy}

\author{Jakub Zakrzewski} 
\email{jakub.zakrzewski@uj.edu.pl}
\affiliation{Instytut Fizyki Teoretycznej, Wydzia\l{} Fizyki, Astronomii i Informatyki Stosowanej,
Uniwersytet Jagiello\'nski,  \L{}ojasiewicza 11, PL-30-348 Krak\'ow, Poland}
\affiliation{Mark Kac Complex Systems Research Center, Uniwersytet Jagiello{\'n}ski, Krak{\'o}w, Poland}

\author{Marcello Dalmonte}
\affiliation{The Abdus Salam International Centre for Theoretical Physics (ICTP), Strada Costiera 11, 34151 Trieste, Italy}
\affiliation{International School for Advanced Studies (SISSA), Via Bonomea 265, I-34136 Trieste, Italy}
\date{\today}

\begin{abstract}
We present a thorough investigation of nonstabilizerness - a fundamental quantum resource that quantifies state complexity within the framework of quantum computing - in a one-dimensional U(1) lattice gauge theory including matter fields. We show how nonstabilizerness is always extensive with volume, and has no direct relation to the presence of critical points. However, its derivatives typically display discontinuities across the latter: This indicates that nonstabilizerness is strongly sensitive to criticality, but in a manner that is very different from entanglement (that, typically, is maximal at the critical point). Our results indicate that error-corrected simulations of lattice gauge theories close to the continuum limit have similar computational costs to those at finite correlation length and provide rigorous lower bounds for quantum resources of such quantum computations. 

\end{abstract}

\maketitle

{\it Introduction. -} There is presently considerable interest in exploring quantum computing and simulation approaches to particle physics phenomena~\cite{Wiese2013,preskill2018simulating,banuls2020simulating,Bauer2023}. A pivotal role in this context is played by lattice gauge theories (LGTs) -  regularized version of field theories defined on discrete space~\cite{wilson1974confinement,montvay1994quantum}. Following early theory proposals~\cite{banerjee2012atomic,zohar2013quantum,zache2018quantum,Surace2020}, this has led to remarkable experiments in analog quantum simulation of abelian LGT in one spatial dimension~\cite{mil2020scalable,yang2020observation,zhou2022thermalization}, at the boundary of classical computational capabilities~\cite{bernien2017probing}. In parallel, digital approaches - where the system dynamics is implemented via a sequence of gates - are also increasingly developing. Early theoretical studies in LGTs demonstrated efficient implementations with the number of qubits~\cite{byrnes2006simulating} and have progressively moved from pioneering demonstrations~\cite{martinez2016real,nguyen2022digital} to mid-scale experiments~\cite{farrell2024scalable,farrell2024quantumsimulationshadrondynamics}. 

With the advent of quantum error correction - whose early steps have been recently demonstrated~\cite{nigg2014quantum,bluvstein2024logical} - digital quantum simulations can be particularly fit for LGTs: The main reason is that their complex multi-body interactions are challenging for analog approaches (describing both minimal coupling and magnetic terms), but do not constitute a major complication for digital ones. 
However, it is presently very unclear what the bottleneck in this direction will be: The key resource in the context of error-correcting codes - nonstabilizerness~\cite{gupta2024encoding} - is essentially unexplored in LGTs, and its potential role in limiting the exploration of gauge theories towards the continuum limit is not known. 

Here, we provide an in-depth study of nonstabilizerness, also known as ``magic'', in the lattice Schwinger model - an archetypal gauge theory in one spatial dimension. Concretely, we utilize the quantum link formalism of LGTs: This formulation of gauge theories utilize spin variables to describe gauge fields, and is the one that has found the widest application so far in the context of quantum simulation \cite{chandrasekharan97,Wiese2013,Aidelsburger2021}. We compute, analytically and numerically, a measure of nonstabilizerness known as stabilizer R\'enyi entropy (SRE)~\cite{Leone2022stabilizer} over the entire phase diagram, which encompasses a variety of phases and (critical) transition lines. Recently, several methods based on Tensor Network states \cite{Banuls2023, ran2020tensor, Schollwoeck2011, Orus2014annphys} have been proposed to compute SRE, utilizing both exact \cite{haug2023quantifying, Tarabunga2024nonstabilizerness} and sampling-like \cite{haug2023stabilizer, Lami2023, tarabunga2023critical, Tarabunga2023many,ballarin2024optimalsamplingtensornetworks} approaches. In this work, we employ a combination of exact diagonalization and Monte Carlo sampling~\cite{tarabunga2023magic}.

\begin{figure}[t]
    \centering
    \includegraphics[width=0.99 \columnwidth]{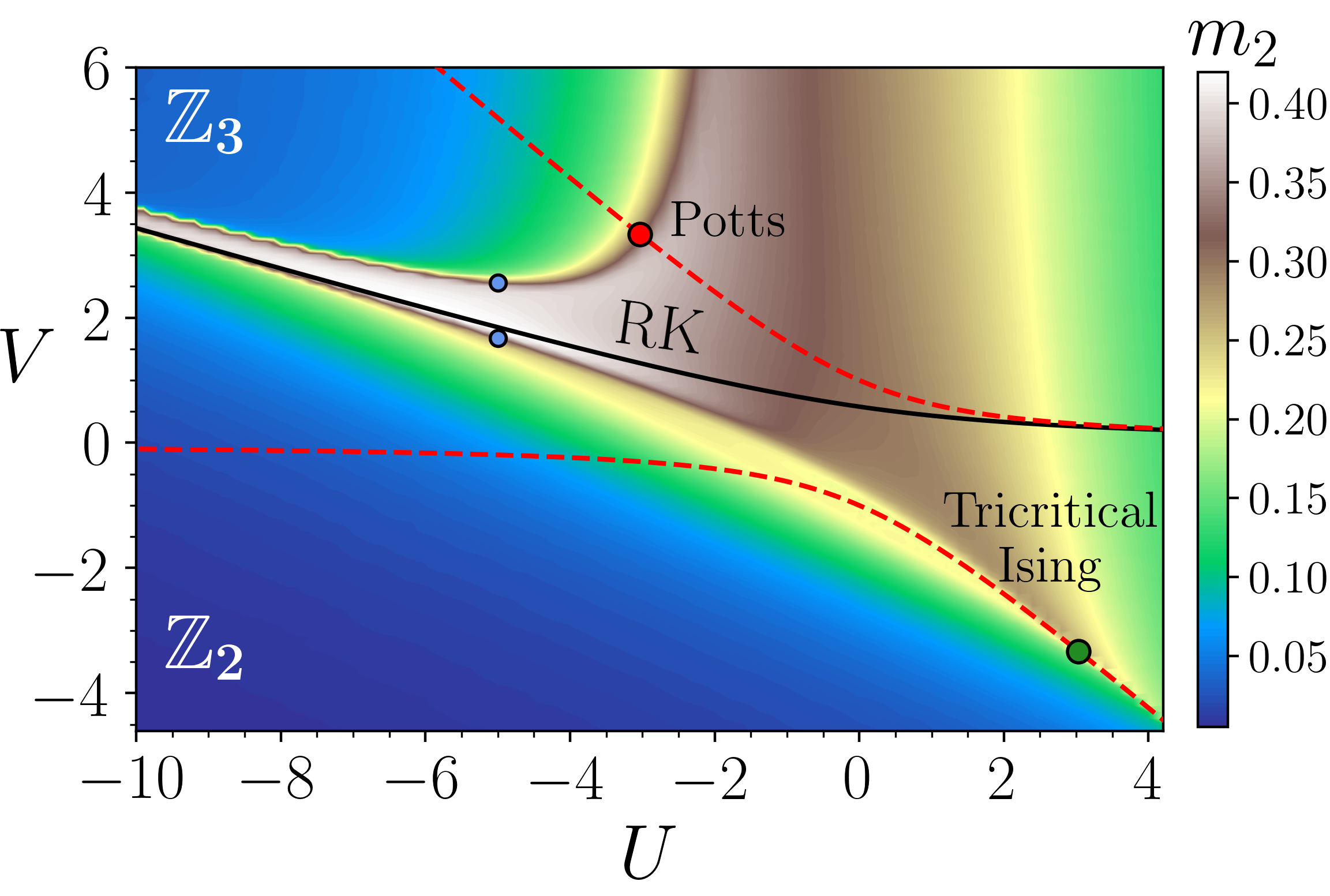}
    \caption{Gauge invariant nonstabilizerness across the phase diagram of the U(1) quantum link model for $L=30$. The black solid line represents an exactly solvable line, \eqref{eq:m2_exact}, while the red dashed lines represent the integrable lines. The red (dark green) circle shows the location of the Potts (tricritical Ising) point in the phase diagram, while the blue points show the location of the maximum derivatives across the transition at $U=-5$.}
    \label{Fig:PDScheme}
\end{figure}

Our results indicate that, regardless of the phase of matter considered, nonstabilizerness is always extensive. However, the magnitude of such extensive scaling varies widely over the phase diagram: Ordered and disordered phases
are characterized by low and high values of nonstabilizerness, respectively, with critical lines in between. Most importantly, our work provides a first throughout investigation of the relation between magic and criticality in gauge theory - a fact that is presently debated even in the context of  spin chains, where different scenarios have been reported~\cite{haug2023stabilizer,tarabunga2023critical,Tarabunga2023many,tarabunga2023magic}. The absolute value of nonstabilizerness fails, in general, to detect transition points - as already observed in other models \cite{Tarabunga2023many,tarabunga2023magic} -, while maxima of derivatives locates critical points precisely \cite{tarabunga2023magic,Tarabunga2024nonstabilizerness,white2021conformal}. 

The aforementioned findings suggest that the digital quantum simulation of gauge theories on error-corrected devices will generically require at least linear with size nonstabilizer resources, whose scaling is, however, not necessarily worst close to the continuum limit. From a basic science viewpoint, the results suggest a strong connection between nonstabilizerness and critical behavior, confirming similar results in the context of statistical mechanics models without gauge symmetries.

{\it Model Hamiltonian. -} The model we are interested in is a lattice version of the Schwinger model~\cite{Schwinger1,Encoding} in the quantum link formulation~\cite{QLink1,QLink2,chandrasekharan97,Brower1999}. It describes the dynamics of fermionic particles, denoted by $\Phi_j$ and residing on the lattice site $j$, mediated by a $U(1)$ gauge field. The latter is represented by a spin-1/2 operator residing on each bond, with the electric field $\hat E_{j,j+1} = S^z_{j,j+1}$. We employ Kogut-Susskind (staggered) fermions~\cite{KogutSusskindFormulation}, with the conventions that holes on odd sites represent antiquarks $\bar q$, while particles on even sites represent quarks $q$. Their  
dynamics is described by~\cite{banerjee2012atomic}:

\begin{eqnarray}\label{eq:QLM}
   H&=&-w\sum_{j=1}^{L}(\Phi^\dag_j S^+_{j,j+1} \Phi^{\mathstrut}_{j+1} + \textnormal{h.c.}) + m\sum_{j=1}^L(-1)^j \Phi^\dag_j \Phi^{\mathstrut}_{j} \label{Schwinger lattice}  \nonumber \\
   &+& V\sum_{j=1}^L S^z_{j-1,j} S^z_{j+1,j+2} + 
   J\sum_{j=1}^{L} (S^z_{j,j+1}-\theta/\pi)^2.    
\end{eqnarray}
Here, the first term describes the minimal coupling between gauge and matter fields, the second is the mass term, the third is a four-Fermi coupling between matter particles, and the last is the electric field, with $\theta$ being a topological angle that, in the following, we will fix to $\theta=\pi$. The generators of the $U(1)$ gauge symmetry are defined as 
\begin{equation}
G_j= E_{j,j+1}-E_{j-1,j}-\Phi^{\dag}_j\Phi^{\mathstrut}_j+\frac{1-(-1)^j}{2},
\label{eq:Gauss}
\end{equation}
and satisfy $[H, G_j] = 0$, so that gauge invariant states $\Ket{\Psi}$ satisfy Gauss law $G_j\Ket{\Psi}=0$ for all values of $j$. 

For practical reasons, it is convenient to work in a formulation where the matter fields are integrated out. The corresponding duality has been derived in Ref. \cite{Surace2020}, showing that Eq.~\eqref{eq:QLM} is equivalent to the Fendley-Sengupta-Sachdev spin-1/2 chain
\cite{fendley2004}:
\begin{equation}
    H = w\sum_{j=1}^{L} \sigma^x_j + U \sum_{j=1}^{L} n_j  + V \sum_{j=1}^{L} n_{j} n_{j+2},
    \label{FSS_H}
\end{equation}
with the constraint $n_j n_{j+1} = 0$. Here, $n_j = (\sigma^z_j+1)/2$, $m=-(U+V)/2$, and we set $w=-1$. 

The model phase diagram, schematically depicted in Fig.~\ref{Fig:PDScheme}, is very rich. It features two symmetry-broken phases (that we denote $\mathbb{Z}_2$ and $\mathbb{Z}_3$ phases, as they break translation symmetry by 2 or 3 lattice sites), a paramagnetic phase, and a gapless phase (whose extent is however very small, and only established at very strong coupling). Starting from $V=0$, the transition separating $\mathbb{Z}_2$ and paramagnetic phase belongs to the Ising universality class and then becomes first order after a tricritical Ising point.  The transition between $\mathbb{Z}_3$ and the paramagnetic phase is debated. However, along the integrable line (red-dashed in Fig.~\ref{Fig:PDScheme}), critical behavior is dictated by the Potts universality class, and it is located precisely at $V_{\mathrm{Potts}} = (\frac{\sqrt{5}+1}{2})^{5/2}$. 

{\it Gauge invariant nonstabilizerness. -} Our main goals are to understand how many non-Clifford resources are needed to represent ground states of $H$ in different phases, and if, when approaching the continuum limit (that is, close to critical lines), nonstabilizerness behaves in a specific manner or not.
In the context of quantum computing and digital quantum simulation, this question can either be asked on the entire Hilbert space, or on the gauge invariant Hilbert space. In terms of resource efficiency, it is natural to disregard the former and focus on the latter case (we note that the latter is also a rigorous upper bound for the former). The reasons are multiple: (a) in the latter case, the number of qubits needed is much smaller with respect to the first case; and (b) by focusing directly on implementing the dynamics within the gauge invariant subspace, issues of gauge violation (which are known to lead to considerable computational overheads, both in terms of postselection and in terms of Hamiltonian engineering) are automatically resolved. We will thus focus on the amount of nonstabilizerness required to realize states using Clifford gates directly onto the gauge invariant subspace: We denote this as gauge invariant nonstabilizerness.

In the dual formulation, the only remnant of the original gauge invariance is a nearest-neighbor constraint: this implies that our nonstabilizerness quantification shall be intended as the amount of non-Clifford resources necessary to generate a state {\it without} leaving the gauge invariant sector of the theory. For the case of global symmetries, an in-depth study of these aspects has been reported in Ref.~\cite{mitsuhashi_clifford_2023}. 

To address this question, we are going to quantify nonstabilizerness via Stabilizer R\'enyi Entropies (SREs) \cite{Leone2022stabilizer,leone2024stabilizer}. For a pure quantum state $|\psi\rangle$, this quantity is defined as

\begin{equation} \label{eq:sre1}
M_n \left( |\psi \rangle \right)= \frac{1}{1-n} \log \left \lbrace \sum_{P \in \mathcal{P}_L} \frac{\langle \psi | P | \psi \rangle^{2n}}{2^L} \right \rbrace \, .
\end{equation}
where $\mathcal{P}_L$ is the group of Pauli string of $L$ qubits.  
We are going to focus on the special case of $n=2$. SREs and other nonstabilizerness measures have already found widespread application in the context of statistical mechanics models, in particular in the context of critical behavior~\cite{haug2023quantifying, tarabunga2023critical,tarabunga2023magic,Tarabunga2023many, Tarabunga2024nonstabilizerness,passarelli2024nonstabilizernesspermutationallyinvariantsystems,liu2024nonequilibriumquantummontecarlo,Oliviero2022,Frau2024}, for quenched gauge theory~\cite{tarabunga2023magic}, and also real-time dynamics~\cite{turkeshi2024magicspreadingrandomquantum,bejan2023dynamical,fux2023entanglement,tarabunga2024magictransitionmeasurementonlycircuits,lópez2024exact,paviglianiti2024estimatingnonstabilizernessdynamicssimulating,lami2024learningstabilizergroupmatrix,mello2024clifford,ahmadi2024mutual}. SREs are also experimentally measurable \cite{Oliviero_2022,haug2023scalable,tirrito2023,Haug2024,niroula2023phase}.

{\it Exactly solvable line. -} 
We start our investigation by analytically obtaining the SRE-2 {\it density} $m_2$ ($M_2/L$) along the exactly solvable line, represented by the solid black line in Fig.~\ref{Fig:PDScheme}. This line is described by the condition: 

\begin{equation}
    3V^2+UV = 1
\end{equation}
with $V>0$, which can be parameterized by
\begin{equation} \label{eq:line}
\begin{split}
    V &=  e^{\beta} \\ U &= e^{-\beta} -3 e^{\beta}.
\end{split}
\end{equation}
The Hamiltonian on this line has the form of the frustration-free Rokhsar-Kivelson-type Hamiltonian \cite{Ardonne2004,Henley2004,Castelnovo2005}, and the ground state is in the disordered phase for any value of $\beta$.
Along this line, the ground state wavefunction can be written as \cite{Ardonne2004,Henley2004,Castelnovo2005}
\begin{equation} \label{eq:gs_line}
    \ket{ \psi(\beta)} = \frac{1}{\sqrt{Z(\beta)}} \sum_{s \in F_N} e^{\beta \sum_i \frac{\sigma^z_i}{2}} \ket{s}
    \, ,\,
    Z(\beta) = \sum_{s \in F_N} e^{\beta \sum_i s_i}
    \, .\nonumber
\end{equation}
Here, $F_N$ is the set of configurations for a periodic chain of $N$ spins, which respects the nearest-neighbor blockade constraint. $Z(\beta)$ can be seen as a classical partition function at temperature $T=1/\beta$. The wavefunction has an MPS representation with bond dimension $\chi=2$ \cite{Okunishi2022}, so bipartite entanglement is bound by $\ln 2$.

We compute $m_2$ along this line using the technique of Ref. \cite{tarabunga2023magic}, yielding
\begin{eqnarray} \label{eq:m2_exact}
m_2 = -\log \frac{16 \lambda}{\left(e^{-\beta} + \sqrt{e^{-2\beta}+4}\right)^4}, 
\end{eqnarray}
for large $L$. Here, $\lambda$ is the largest root of the cubic equation
\begin{equation}
    x^3 - (1+e^{-4 \beta}) x^2 - (1+13e^{-4 \beta}) x +1 = 0.
\end{equation}
The corresponding result is depicted in Fig.~\ref{Fig:RK_line}(a), and indicates that {\it (i)} $M_2$ is always extensive along the entire solvable line; and {\it (ii)} gauge invariant nonstabilizerness has a very distinct behavior with respect to entanglement, that is instead size-independent and almost constant over the entire solvable line. It is instead not immediately clear how nonstabilizerness and critical behavior are related (the regime closest to criticality is for $V\gtrsim 3$): To address this point, it is important to move away from the solvable line and investigate the phase diagram using numerical simulations.

\begin{figure}[t!]
    \centering
    \includegraphics[width=0.86 \columnwidth]{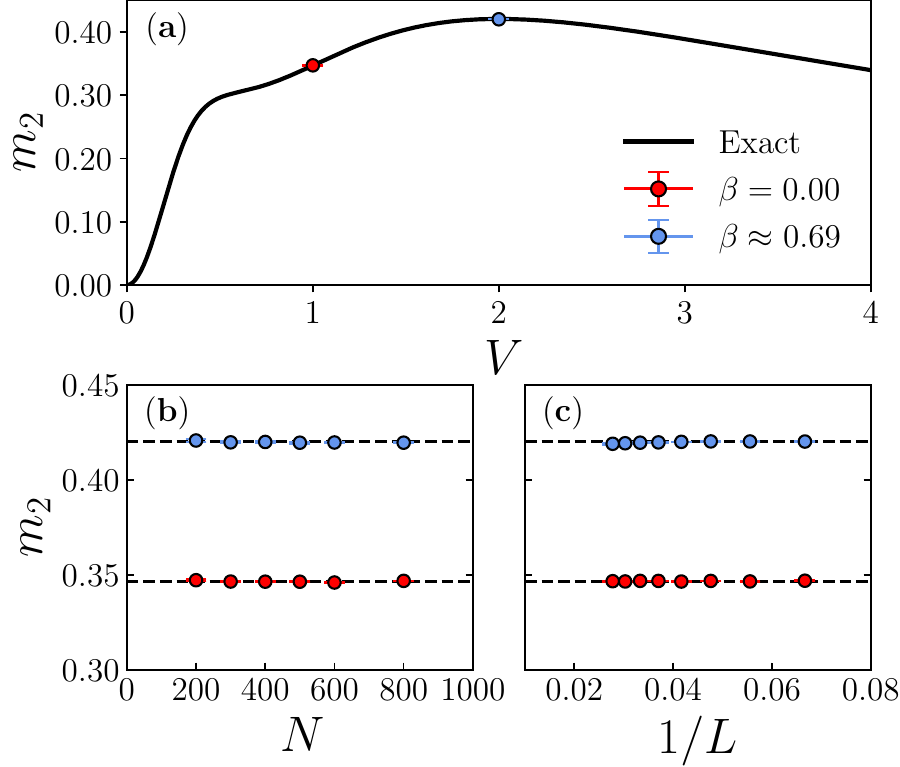}
    \caption{(a) SRE-2 density versus $V$ in the RK line. The solid line represents the exact result obtained using \eqref{eq:m2_exact}, while red (blue) circles are the numerical result for $\beta=0$ ($\beta \approx 0.69$) with $L=30$; (b) Numerical convergence of $m_2$ with the number of sampled states $N$ for two different points in the solvable line. For $N=800$, the numerical error in $m_2$ is less than $1\%$. (c) Finite-size effects in $m_2$ for two different points in the RK line.}
    \label{Fig:RK_line}
\end{figure}

{\it Towards the continuum limit: Nonstabilizerness and critical points. -}
We obtain the ground state of \eqref{FSS_H} via exact diagonalization (ED) of periodic chains. The combination of gauge invariance and translational invariance significantly reduces the Hilbert space dimension, allowing us to obtain and sample the ground state of systems up to $42$ spins~\cite{supmat}.

Evaluating \eqref{eq:sre1} using methods such as the Pauli-Markov chain \cite{Tarabunga2023many,tarabunga2023critical} becomes challenging due to the large number of samples required to achieve relatively small errors, while via direct sampling is prohibitive due to precision issues. However, for any state $|\psi \rangle = \sum_s c_s|s\rangle$, the SRE-2 can be rewritten in a more convenient expression for Monte Carlo sampling \cite{tarabunga2023magic}

\begin{equation} \label{eq:sre2}
    \begin{split}
    \exp(-M_2) = 
    \!\!\!\!\sum_{s^{(1)},s^{(2)},s^{(3)},s^{(4)}}\! 
    \left[ \vphantom{\sum}
    c_{s^{(1)}} c_{s^{(2)}} c_{s^{(3)}}  c_{s^{(1)}s^{(2)}s^{(3)}} 
    \right.
    \\
    \left. \vphantom{\sum}
    c^{*}_{s^{(1)} s^{(2)} s^{(4)}} c^{*}_{s^{(1)} s^{(3)} s^{(4)}} c^{*}_{s^{(2)}s^{(3)}s^{(4)}} 
    c^{*}_{s^{(4)}} 
    \right] \, , 
    \end{split}
\end{equation}
where $s^{(a)}$ represents the basis states. 
This formulation allows direct sampling over the ground state coefficients $|c_s|^2$, enabling perfect sampling by considering $\binom{N}{4}$ combinations from $N$ samples. To validate this method - applied to gauge theories for the first time - we confirmed that $m_2$ matches the analytical solution from Eq.~\ref{eq:m2_exact} for two different points along the solvable line, depicted as red and blue circles in Fig.~\ref{Fig:RK_line}. It yields small errors for different values of $N$ and no significant finite-size effects are observed; see~\cite{supmat} for more details.

\begin{figure}[]
    \centering
    \includegraphics[width=0.86 \columnwidth]{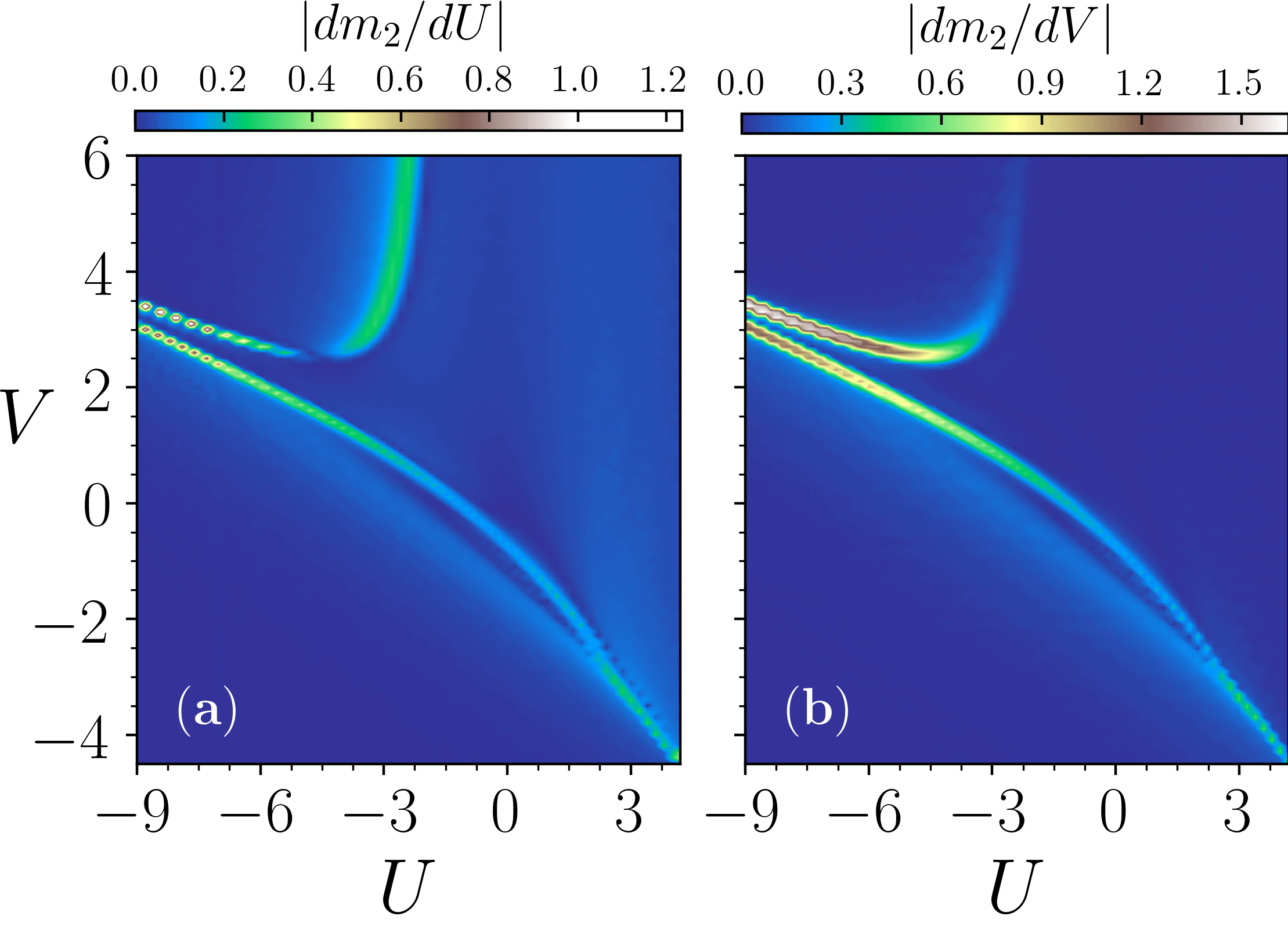}
    \caption{Absolute value of the first derivative of the SRE-2 density with respect to (a) $U$ and (b) $V$ across the entire phase diagram (Fig.~\ref{Fig:PDScheme}). The transition lines that separate ordered-disordered phases are well captured by the first derivatives, showing the sensitivity of nonstabilizerness to criticality. }
    \label{Fig:Derivatives_PD}
\end{figure}

We now move away from the solvable line and explore the relation between gauge invariant nonstabilizerness and phase transitions. In Fig.~\ref{Fig:PDScheme}, we plot $m_2$ across three main phases of the model. In ordered phases, the SRE-2 density has low values of nonstabilizerness, with its magnitude decreasing as we delve deeper into these phases. Oppositely, $m_2$ has significantly higher values in the disordered phase and it behaves nonuniformly across it. Its maximum happens near the exact solvable line, reaching $m_2 \approx 0.42$.

With $m_2$ showing a different behavior across different phases, it seems natural to look at the first derivatives of $m_2$. 
In Fig.~\ref{Fig:Derivatives_PD}, we show how the derivatives capture the ordered-disordered transition across the FSS diagram, with the transition lines between different phases being well delimited. Therefore, similarly to entanglement, gauge-invariant nonstabilizerness is sensitive to criticality, but in a distinct manner: Instead of being maximal at the critical point, the phase transitions are signaled by a maximum of the first derivatives of $m_2$, which shows discontinuities at these points. Let us note that since the wavefunction is not analytic at the critical point, observables or their derivatives should reveal singularities - see, e.g., the Bose-Hubbard model~\cite{Lacki2016,Lacki2021}.

\begin{figure*}[]
    \centering
    \includegraphics[width=0.82\linewidth]{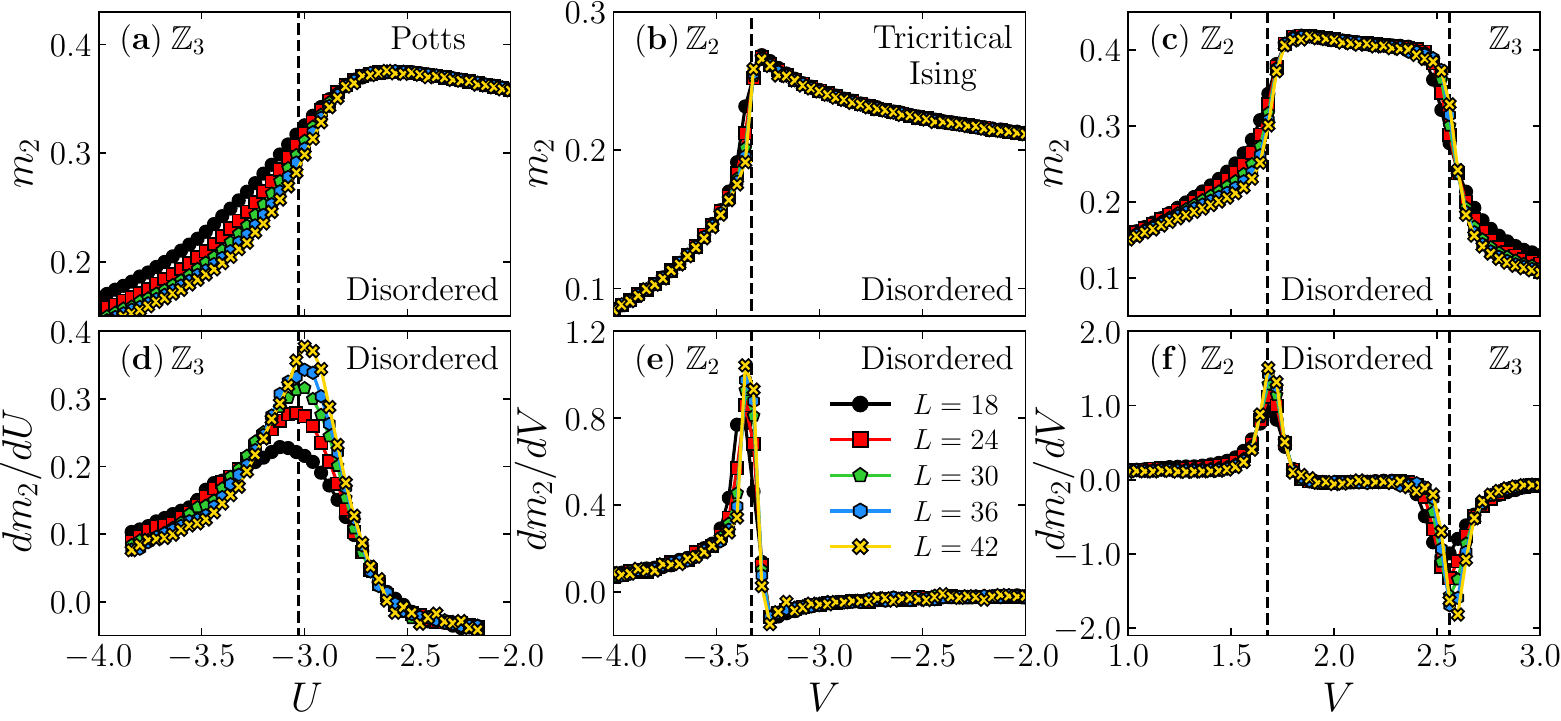}
    \caption{SRE-2 density $m_2$ (upper panel) and its derivative (lower panel) at various cuts in the phase diagram. The critical points are determined by the expression $U=-V + 1/V$, with the Potts (tricritical Ising) critical point occurring at $V = V_\mathrm{Potts}$ (or $V = -V_\mathrm{Potts})$. The first derivatives peak precisely at these critical points, as shown in panels (d) and (e). Panels (c) and (f) illustrate the behavior of $m_2$ and its first derivative at $U=-5.0$, respectively. }
    \label{Fig:Magic_Cuts}
\end{figure*}

{\it Potts Critical point}: We now clarify the connection between nonstabilizerness and field theories, taking place along critical lines. We start by inspecting the case $V_{\mathrm{Potts}} = (\frac{\sqrt{5}+1}{2})^{5/2}$, examining how $m_2$ behaves as a function of $U$, as shown in Fig.~\ref{Fig:Magic_Cuts}(a). It shows a smooth transition from $\mathbb{Z}_3$ ordered phase, where $m_2$ has relatively low values as displays a sublinear scaling with $L$, to a perfect linear scaling in the disordered phase, which does not show any finite-size effect in the SRE-2 density $m_2$~\cite{supmat}. There is no indication of phase transitions at the critical point, represented here as a dashed line. However, as depicted in Fig.~\ref{Fig:Magic_Cuts}(d), the first derivative of $m_2$ peaks precisely at the critical point, showing the sensitivity of the first derivative to the criticality. 

{\it Tricritical Ising point}: 
In the vicinity of the tricritical Ising point, the scaling of $m_2$ is different, as it seems to peak in the vicinity of the transition (where it also becomes size-independent). Most importantly, however, its derivative still peaks at the critical point, exhibiting a size-dependent maximum. 

{\it Phase diagram cut at $U=-5$}: We then check how $m_2$ behaves across a phase diagram cut, featuring two ordered phases separated by a disordered one [see Fig.~\ref{Fig:Magic_Cuts}(c)]. As expected, the value of $m_2$ is maximum in the disorder phase and does not show very clear features at transition points. Instead, again, its derivative displays two clear peaks at the transition points. These two different points in $V$ correspond to the blue points in the phase diagram (Fig.~\ref{Fig:PDScheme}), showing that indeed the location of the disordered phase can be captured by the first derivative of $m_2$. We thus conclude that, for the model at hand, while the value of $m_2$ does not behave generically close to the continuum limit, its derivative always shows signs of critical behavior~\footnote{The sign of the derivative depends on which direction the transition is approached and is thus non-generic.}.

{\it Conclusions and outlook. -} We have investigated how nonstabilizerness behaves in lattice gauge theory, introducing the notion of gauge invariant nonstabilizerness - that is, nonstabilizerness in the context of gauge invariant operations. In 1D U(1) lattice gauge theory, full-state nonstabilizerness peaks in trivial phases but does not necessarily indicate the presence of phase transitions, while its derivatives always do. This signals that nonstabilizerness, in gauge theories, behaves remarkably differently from entanglement. From a practical viewpoint, our results indicate that realizing gauge theory dynamics on error-corrected quantum computers might be particularly resource intensive, as, in the continuum limit, nonstabilizerness is always extensive. The fact that our lower bounds are nevertheless relatively small shall motivate further study on optimal compilation, which, so far, have returned very major resource requirements~\cite{Lamm19,Murairi22,ciavarella2022preparation,Davoudi23}. It would be interesting to explore such features for non-Abelian lattice gauge theories, where analog and digital implementations have gathered significant attention over the past years - some of them focusing precisely on the model we discuss here~\cite{Davoudi23,osterloh2005cold,banerjee2013atomic,tagliacozzo2013simulation,mezzacapo2015non,zohar2017digital,davoudi2021search, calajo2024digital}, 
as well as within effective models relevant for particle theory~\cite{robin2024magic,white2024magic}.

{\it Acknowledgments. -}We thank M. Collura, T. Haug and P. Hauke for discussions, and A. Paviglianiti for comments on the manuscript. P.R.N.F thanks A.S. Aramthottil and Konrad Pawlik for the valuable discussion on the numerics. Discussions with B. Damski are also appreciated (J.Z.). E.T. acknowledges J. Mildenberger, G. Chandra Santra, and R. Costa de Almeida for insightful discussions.
We acknowledge Polish high-performance computing infrastructure PLGrid for awarding this project access to the LUMI supercomputer, owned by the EuroHPC Joint Undertaking, hosted by CSC (Finland) and the LUMI consortium through PLL/2023/04/016502.
The work of P.R.N.F. and that of J.Z. was
funded by the National Science Centre, Poland, project
2021/03/Y/ST2/00186 within the QuantERA II Programme (DYNAMITE) that has received funding from the European
Union Horizon 2020 research and innovation programme
under Grant agreement No 101017733.  The study was also partially funded by the "Research Support Module" as part of the "Excellence Initiative – Research University" program at the Jagiellonian University in Kraków.
M.\,D.and E. T. were partly supported by the QUANTERA DYNAMITE PCI2022-132919. M.\,D. was also supported by the EU-Flagship programme Pasquans2, by the PNRR MUR project PE0000023-NQSTI, the PRIN programme (project CoQuS), and the ERC Consolidator grant WaveNets. P. S. T. acknowledges support from the Simons Foundation through Award 284558FY19 to the ICTP. 
Views and
opinions expressed in this work are, however, those of the
authors only and do not necessarily reflect those of the
European Union, European Climate, Infrastructure and
Environment Executive Agency (CINEA), nor any other
granting authority

{\it Note added:} While finalizing this work, we became aware of a study of stabilizer Renyi entropies in the context of PXP models~\cite{smith2024nonstabilizerness}. The two works utilize very different techniques to compute SRE-2 density $m_2$ and are complementary in terms of parameter regimes. In the only overlapping computation (discussed in Ref.~\cite{supmat}), the agreement between out analytics and the simulations in Ref.~\cite{smith2024nonstabilizerness} is perfect. 

%


\newcommand{\snum}{S}

\setcounter{equation}{0}
\setcounter{figure}{0}
\renewcommand{\thefigure}{S-\arabic{figure}}
\setcounter{table}{0}
\setcounter{page}{1}
\renewcommand{\thepage}{S\arabic{figure}}

\renewcommand{\theequation}{\snum.\arabic{equation}}

\maketitle

\section*{Supplementary material: 
 Nonstabilizerness in U(1) lattice gauge theory }

\subsection{Nonstabilizerness of “rainbow”-like state}
Here, we show that our results can also be used to compute the SRE of the “rainbow”-like state
\begin{equation}
    \ket{E} = \frac{1}{\sqrt{\chi}} \sum_{f \in F_L} (-1)^{|f|}  \ket{f} \otimes \ket{f} ,
\end{equation}
which is an exact volume-law entangled eigenstate at energy $E = 0$ in the PXP chain with PBCs and size $N = 2L$. Here, $|f|$ denotes the parity of a given configuration. The normalization is given by $\chi=Z_{L}(\beta=0)$. We first notice that applying the operator $\otimes_{i=1}^L \sigma^z_i$ to $\ket{E}$ removes the $(-1)^{|f|}$ factor. Since the operator is a Clifford gate, the SRE-2 is unchanged. We immediately see that the SRE density of $\ket{E}$ is given by half of the SRE density of $\ket{\psi(\beta=0)}$, which is $m_2 \approx 0.1733$. This result is within the errorbar of the numerical calculation in Ref. \cite{smith2024nonstabilizerness}. This highlights the wide applicability of the method employed here, as it is useful beyond states efficiently represented by low bond dimension tensor networks. 

\subsection{Nonstabilizerness is always extensive}

\begin{figure}[h!]
    \centering
    \includegraphics[width=1 \columnwidth]{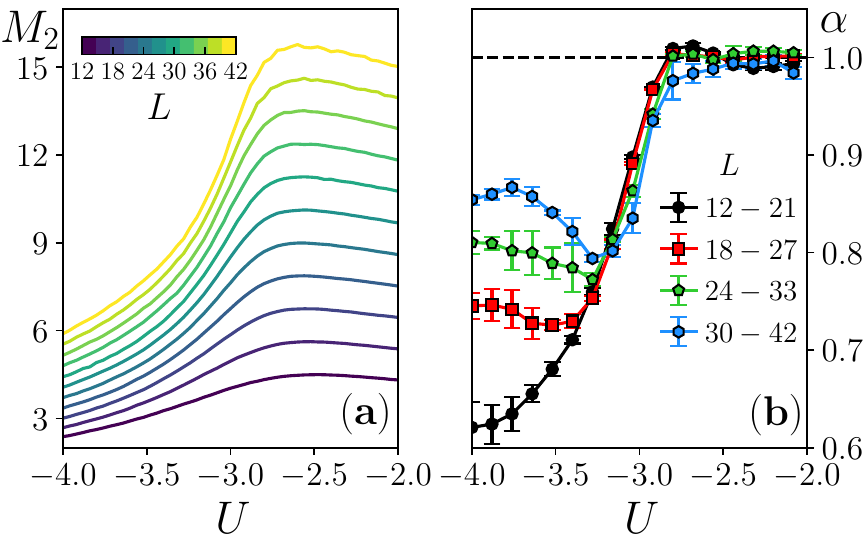}

    \caption{(a) System-size dependence of $M_2$ near the Potts critical point, showing that the gauge-invariant nonstabilizerness is always an extensive quantity; (b) Power-law exponent $\alpha$ at different fitting ranges. Although a sublinear behavior is observed in the ordered phase, it flows towards a linear scaling in the {\it asymptotic} limit. }
    \label{Fig:Potts}
\end{figure}

In this section, we demonstrate that $M_2$ is always an extensive quantity, even in ordered phases. In Fig.~\ref{Fig:Potts}(a), we can see that $M_2$ is always extensive with system-size $L$ near the Potts critical point and, therefore, the amount of non-Clifford resources needed to simulate the two distinguish phases always grows with $L$. 

However, one can observe that $M_2$ has a sublinear dependence with $L$, as it is clear from Fig. 4(a) in the main text. This behavior is merely a finite-size effect, with the asymptotic scaling being a linear dependence with $L$. To illustrate this, we assume a power-law growth of $M_2$ ($M_2 = AL^{\alpha}$) and show in Fig.\ref{Fig:Potts}(b) how the power-law exponent $\alpha$ changes assuming different fitting ranges. In the ordered $\mathbb{Z}_3$ phase, it can be seen that the exponent grows as the system-size range of the fitting increases, and this suggests that the asymptotic behavior of $M_2$ always grows linearly with $L$.

The results in the other cuts analyzed throughout this paper are similar. Near the Tricritical Ising point, the SRE density $m_2$ becomes system size independent, showing that the gauge-invariant nonstabilizerness is a quantity that always grows linearly with system size, independent of the nature of the phase. The same conclusion can be obtained by looking at the third cut in the phase diagram where, despite some finite-size effects at small values of $L$, the curves of $m_2$ for $L=30$,$36$ and $42$ nearly overlap with each other. 

\subsection{Sampling method and error analysis}

The ground state $|\psi\rangle = \sum_s c_s|s \rangle$ of the U(1) QLM is obtained using exact diagonalization directly in its dual formulation (Eq. 3 of the main text). The Gauss law (nearest-neighbor constraint in the dual formulation) strongly reduces the growth of the Hilbert space dimension, which scales as $\mathrm{dim}\mathcal{H}_{L} = F_{L-1} + F_{L+1}$, where $F_{i}$ is the $i$-th Fibonacci number and $L$ is the number of matter sites. In the limit of large $L$, the Hilbert space dimension is given by $\mathrm{dim}\mathcal{H}_{L} \sim (1.618)^L$. Moreover, the periodic boundary conditions allow us to diagonalize the system directly within the momentum sector where the ground state resides. Therefore, by leveraging these two features, we can obtain the ground state of systems up to $L=42$ matter sites using Krylov space techniques.

Once we obtain the ground state in the momentum space, we transform it back to the real space. We then perform perfect sampling directly on its coefficients $|c_s|^2$ and compute all $\binom{N}{4}$ possibilities to choose from $N$ samples selected according to the cumulative distribution of the ground state coefficients (see Eq. 9 of the main text). 

Although the gauge constraints eliminate many possibilities, we achieve a good numerical convergence by setting a sufficiently large number of samples and averaging over $50$ different realizations  (see details below). The precision of our method is shown in Fig. 2(a) of the main text, where we benchmark the numerical results along the exact solvable line. It shows perfect agreement with the analytical prediction (black solid line), illustrating the robustness of our numerical algorithm. The convergence of $m_2$ with $N$ is relatively fast for moderate system sizes ($L=30$ qubits), as shown in Fig. 2(b), yielding errors of less than $1\%$ for any value of $N$ considered. Additionally, no finite-size effects are observed, with numerics matching the analytical prediction already for systems with $L=15$ spins (c.f. Fig. 2(c)).

\begin{figure}[t!]
    \centering
    \includegraphics[width=0.65\linewidth]{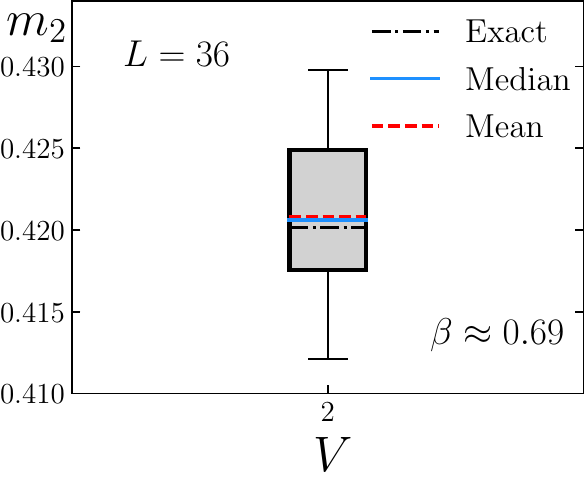}
    \caption{Boxplot of $50$ different runs along the exact solvable line ($V=2$) for $L=36$ qubits. Although different simulations yield different results, the difference between the mean value (red dashed line) and the exact solution (black dashed-dotted line) is less than $0.2\%$, showing the accuracy of our results.}
    \label{Fig:Appendix}
\end{figure}

It is important to stress that, because of the gauge constraint, a single Monte Carlo run may not be sufficient to obtain well-converged results. Therefore, to minimize the uncertainties in the computation of $m_2$, we perform $50$ independent runs and average the results. The statistical uncertainty is obtained by assuming that different runs are uncorrelated with each other, and is given by

\begin{equation}
    \sigma_{m_2} = \sqrt{\frac{\langle (m_2 - \langle m_2\rangle)^2 \rangle}{N_{\mathrm{real}}}}
\end{equation}
where $ \langle m_2 \rangle$ is the average over $ N_\mathrm{real}$ different Monte Carlo runs. To benchmark the convergence of our numerical procedure, we set $N=500$ and evaluate the algorithm at a specific point along the solvable line (Eq. 7 of the main text) where the system has a large amount of nonstabilizerness ($m_2 \approx 0.42$). The different results for $L=36$ are illustrated in Fig.~\ref{Fig:Appendix} in a boxplot representation, which conveniently shows how spread different runs can be. The difference between different runs is due to the Gauss law, which highly constrains the model so that the number of nonzero elements in the sum is much less than $\binom{N}{4}$ terms. However, the relative error between the mean value (red dashed lines) and the exact result (black dashed-dotted line) is approximately $0.16\%$, showing the robustness of our results.

It is important to mention that, throughout the letter, we consider at least $N=500$ samples. Therefore, despite the constraints imposed by the Gauss law, the statistical uncertainty is less than $1\%$, even for the largest system size available ($L=42$). However, our method loses its precision very quickly as $L$ grows, and one needs to consider $N>500$ samples for higher system sizes.


\begin{thebibliography}{87}%
\makeatletter
\providecommand \@ifxundefined [1]{%
 \@ifx{#1\undefined}
}%
\providecommand \@ifnum [1]{%
 \ifnum #1\expandafter \@firstoftwo
 \else \expandafter \@secondoftwo
 \fi
}%
\providecommand \@ifx [1]{%
 \ifx #1\expandafter \@firstoftwo
 \else \expandafter \@secondoftwo
 \fi
}%
\providecommand \natexlab [1]{#1}%
\providecommand \enquote  [1]{``#1''}%
\providecommand \bibnamefont  [1]{#1}%
\providecommand \bibfnamefont [1]{#1}%
\providecommand \citenamefont [1]{#1}%
\providecommand \href@noop [0]{\@secondoftwo}%
\providecommand \href [0]{\begingroup \@sanitize@url \@href}%
\providecommand \@href[1]{\@@startlink{#1}\@@href}%
\providecommand \@@href[1]{\endgroup#1\@@endlink}%
\providecommand \@sanitize@url [0]{\catcode `\\12\catcode `\$12\catcode `\&12\catcode `\#12\catcode `\^12\catcode `\_12\catcode `\%12\relax}%
\providecommand \@@startlink[1]{}%
\providecommand \@@endlink[0]{}%
\providecommand \url  [0]{\begingroup\@sanitize@url \@url }%
\providecommand \@url [1]{\endgroup\@href {#1}{\urlprefix }}%
\providecommand \urlprefix  [0]{URL }%
\providecommand \Eprint [0]{\href }%
\providecommand \doibase [0]{https://doi.org/}%
\providecommand \selectlanguage [0]{\@gobble}%
\providecommand \bibinfo  [0]{\@secondoftwo}%
\providecommand \bibfield  [0]{\@secondoftwo}%
\providecommand \translation [1]{[#1]}%
\providecommand \BibitemOpen [0]{}%
\providecommand \bibitemStop [0]{}%
\providecommand \bibitemNoStop [0]{.\EOS\space}%
\providecommand \EOS [0]{\spacefactor3000\relax}%
\providecommand \BibitemShut  [1]{\csname bibitem#1\endcsname}%
\let\auto@bib@innerbib\@empty
\bibitem [{\citenamefont {Wiese}(2013)}]{Wiese2013}%
  \BibitemOpen
  \bibfield  {author} {\bibinfo {author} {\bibfnamefont {U.~J.}\ \bibnamefont {Wiese}},\ }\bibfield  {title} {\bibinfo {title} {{Ultracold quantum gases and lattice systems: quantum simulation of lattice gauge theories}},\ }\href {https://doi.org/10.1002/andp.201300104} {\bibfield  {journal} {\bibinfo  {journal} {Annalen der Physik}\ }\textbf {\bibinfo {volume} {525}},\ \bibinfo {pages} {777} (\bibinfo {year} {2013})}\BibitemShut {NoStop}%
\bibitem [{\citenamefont {Preskill}(2018)}]{preskill2018simulating}%
  \BibitemOpen
  \bibfield  {author} {\bibinfo {author} {\bibfnamefont {J.}~\bibnamefont {Preskill}},\ }\href@noop {} {\bibinfo {title} {{Simulating quantum field theory with a quantum computer}}} (\bibinfo {year} {2018}),\ \Eprint {https://arxiv.org/abs/1811.10085} {arXiv:1811.10085 [hep-lat]} \BibitemShut {NoStop}%
\bibitem [{\citenamefont {Banuls}\ \emph {et~al.}(2020)\citenamefont {Banuls}, \citenamefont {Blatt}, \citenamefont {Catani}, \citenamefont {Celi}, \citenamefont {Cirac}, \citenamefont {Dalmonte}, \citenamefont {Fallani}, \citenamefont {Jansen}, \citenamefont {Lewenstein}, \citenamefont {Montangero} \emph {et~al.}}]{banuls2020simulating}%
  \BibitemOpen
  \bibfield  {author} {\bibinfo {author} {\bibfnamefont {M.~C.}\ \bibnamefont {Banuls}}, \bibinfo {author} {\bibfnamefont {R.}~\bibnamefont {Blatt}}, \bibinfo {author} {\bibfnamefont {J.}~\bibnamefont {Catani}}, \bibinfo {author} {\bibfnamefont {A.}~\bibnamefont {Celi}}, \bibinfo {author} {\bibfnamefont {J.~I.}\ \bibnamefont {Cirac}}, \bibinfo {author} {\bibfnamefont {M.}~\bibnamefont {Dalmonte}}, \bibinfo {author} {\bibfnamefont {L.}~\bibnamefont {Fallani}}, \bibinfo {author} {\bibfnamefont {K.}~\bibnamefont {Jansen}}, \bibinfo {author} {\bibfnamefont {M.}~\bibnamefont {Lewenstein}}, \bibinfo {author} {\bibfnamefont {S.}~\bibnamefont {Montangero}}, \emph {et~al.},\ }\bibfield  {title} {\bibinfo {title} {{Simulating lattice gauge theories within quantum technologies}},\ }\href {https://doi.org/10.1140/epjd/e2020-100571-8} {\bibfield  {journal} {\bibinfo  {journal} {The European Physical Journal D}\ }\textbf {\bibinfo {volume} {74}},\ \bibinfo {pages} {1} (\bibinfo {year} {2020})}\BibitemShut {NoStop}%
\bibitem [{\citenamefont {Bauer}\ \emph {et~al.}(2023)\citenamefont {Bauer}, \citenamefont {Davoudi}, \citenamefont {Balantekin}, \citenamefont {Bhattacharya}, \citenamefont {Carena}, \citenamefont {De~Jong}, \citenamefont {Draper}, \citenamefont {El-Khadra}, \citenamefont {Gemelke}, \citenamefont {Hanada} \emph {et~al.}}]{Bauer2023}%
  \BibitemOpen
  \bibfield  {author} {\bibinfo {author} {\bibfnamefont {C.~W.}\ \bibnamefont {Bauer}}, \bibinfo {author} {\bibfnamefont {Z.}~\bibnamefont {Davoudi}}, \bibinfo {author} {\bibfnamefont {A.~B.}\ \bibnamefont {Balantekin}}, \bibinfo {author} {\bibfnamefont {T.}~\bibnamefont {Bhattacharya}}, \bibinfo {author} {\bibfnamefont {M.}~\bibnamefont {Carena}}, \bibinfo {author} {\bibfnamefont {W.~A.}\ \bibnamefont {De~Jong}}, \bibinfo {author} {\bibfnamefont {P.}~\bibnamefont {Draper}}, \bibinfo {author} {\bibfnamefont {A.}~\bibnamefont {El-Khadra}}, \bibinfo {author} {\bibfnamefont {N.}~\bibnamefont {Gemelke}}, \bibinfo {author} {\bibfnamefont {M.}~\bibnamefont {Hanada}}, \emph {et~al.},\ }\bibfield  {title} {\bibinfo {title} {Quantum simulation for high-energy physics},\ }\href {https://doi.org/10.1103/PRXQuantum.4.027001} {\bibfield  {journal} {\bibinfo  {journal} {PRX Quantum}\ }\textbf {\bibinfo {volume} {4}},\ \bibinfo {pages} {027001} (\bibinfo {year} {2023})}\BibitemShut {NoStop}%
\bibitem [{\citenamefont {Wilson}(1974)}]{wilson1974confinement}%
  \BibitemOpen
  \bibfield  {author} {\bibinfo {author} {\bibfnamefont {K.~G.}\ \bibnamefont {Wilson}},\ }\bibfield  {title} {\bibinfo {title} {{Confinement of quarks}},\ }\href {https://doi.org/10.1103/PhysRevD.10.2445} {\bibfield  {journal} {\bibinfo  {journal} {Phys. Rev. D}\ }\textbf {\bibinfo {volume} {10}},\ \bibinfo {pages} {2445} (\bibinfo {year} {1974})}\BibitemShut {NoStop}%
\bibitem [{\citenamefont {Montvay}\ and\ \citenamefont {M{\"u}nster}(1994)}]{montvay1994quantum}%
  \BibitemOpen
  \bibfield  {author} {\bibinfo {author} {\bibfnamefont {I.}~\bibnamefont {Montvay}}\ and\ \bibinfo {author} {\bibfnamefont {G.}~\bibnamefont {M{\"u}nster}},\ }\href@noop {} {\emph {\bibinfo {title} {{Quantum fields on a lattice}}}}\ (\bibinfo  {publisher} {Cambridge University Press},\ \bibinfo {year} {1994})\BibitemShut {NoStop}%
\bibitem [{\citenamefont {Banerjee}\ \emph {et~al.}(2012)\citenamefont {Banerjee}, \citenamefont {Dalmonte}, \citenamefont {M{\"u}ller}, \citenamefont {Rico}, \citenamefont {Stebler}, \citenamefont {Wiese},\ and\ \citenamefont {Zoller}}]{banerjee2012atomic}%
  \BibitemOpen
  \bibfield  {author} {\bibinfo {author} {\bibfnamefont {D.}~\bibnamefont {Banerjee}}, \bibinfo {author} {\bibfnamefont {M.}~\bibnamefont {Dalmonte}}, \bibinfo {author} {\bibfnamefont {M.}~\bibnamefont {M{\"u}ller}}, \bibinfo {author} {\bibfnamefont {E.}~\bibnamefont {Rico}}, \bibinfo {author} {\bibfnamefont {P.}~\bibnamefont {Stebler}}, \bibinfo {author} {\bibfnamefont {U.-J.}\ \bibnamefont {Wiese}},\ and\ \bibinfo {author} {\bibfnamefont {P.}~\bibnamefont {Zoller}},\ }\bibfield  {title} {\bibinfo {title} {Atomic quantum simulation of dynamical gauge fields coupled to fermionic matter: From string breaking to evolution after a quench},\ }\href {https://doi.org/10.1103/PhysRevLett.109.175302} {\bibfield  {journal} {\bibinfo  {journal} {Phys. Rev. Lett.}\ }\textbf {\bibinfo {volume} {109}},\ \bibinfo {pages} {175302} (\bibinfo {year} {2012})}\BibitemShut {NoStop}%
\bibitem [{\citenamefont {Zohar}\ \emph {et~al.}(2013)\citenamefont {Zohar}, \citenamefont {Cirac},\ and\ \citenamefont {Reznik}}]{zohar2013quantum}%
  \BibitemOpen
  \bibfield  {author} {\bibinfo {author} {\bibfnamefont {E.}~\bibnamefont {Zohar}}, \bibinfo {author} {\bibfnamefont {J.~I.}\ \bibnamefont {Cirac}},\ and\ \bibinfo {author} {\bibfnamefont {B.}~\bibnamefont {Reznik}},\ }\bibfield  {title} {\bibinfo {title} {{Quantum simulations of gauge theories with ultracold atoms: Local gauge invariance from angular-momentum conservation}},\ }\href {https://doi.org/10.1103/PhysRevA.88.023617} {\bibfield  {journal} {\bibinfo  {journal} {Phys. Rev. A}\ }\textbf {\bibinfo {volume} {88}},\ \bibinfo {pages} {023617} (\bibinfo {year} {2013})}\BibitemShut {NoStop}%
\bibitem [{\citenamefont {Zache}\ \emph {et~al.}(2018)\citenamefont {Zache}, \citenamefont {Hebenstreit}, \citenamefont {Jendrzejewski}, \citenamefont {Oberthaler}, \citenamefont {Berges},\ and\ \citenamefont {Hauke}}]{zache2018quantum}%
  \BibitemOpen
  \bibfield  {author} {\bibinfo {author} {\bibfnamefont {T.~V.}\ \bibnamefont {Zache}}, \bibinfo {author} {\bibfnamefont {F.}~\bibnamefont {Hebenstreit}}, \bibinfo {author} {\bibfnamefont {F.}~\bibnamefont {Jendrzejewski}}, \bibinfo {author} {\bibfnamefont {M.~K.}\ \bibnamefont {Oberthaler}}, \bibinfo {author} {\bibfnamefont {J.}~\bibnamefont {Berges}},\ and\ \bibinfo {author} {\bibfnamefont {P.}~\bibnamefont {Hauke}},\ }\bibfield  {title} {\bibinfo {title} {{Quantum simulation of lattice gauge theories using Wilson fermions}},\ }\href {https://doi.org/10.1088/2058-9565/aac33b} {\bibfield  {journal} {\bibinfo  {journal} {Quantum Science and Technology}\ }\textbf {\bibinfo {volume} {3}},\ \bibinfo {pages} {034010} (\bibinfo {year} {2018})}\BibitemShut {NoStop}%
\bibitem [{\citenamefont {Surace}\ \emph {et~al.}(2020)\citenamefont {Surace}, \citenamefont {Mazza}, \citenamefont {Giudici}, \citenamefont {Lerose}, \citenamefont {Gambassi},\ and\ \citenamefont {Dalmonte}}]{Surace2020}%
  \BibitemOpen
  \bibfield  {author} {\bibinfo {author} {\bibfnamefont {F.~M.}\ \bibnamefont {Surace}}, \bibinfo {author} {\bibfnamefont {P.~P.}\ \bibnamefont {Mazza}}, \bibinfo {author} {\bibfnamefont {G.}~\bibnamefont {Giudici}}, \bibinfo {author} {\bibfnamefont {A.}~\bibnamefont {Lerose}}, \bibinfo {author} {\bibfnamefont {A.}~\bibnamefont {Gambassi}},\ and\ \bibinfo {author} {\bibfnamefont {M.}~\bibnamefont {Dalmonte}},\ }\bibfield  {title} {\bibinfo {title} {{Lattice Gauge Theories and String Dynamics in {Rydberg} Atom Quantum Simulators}},\ }\href {https://doi.org/10.1103/PhysRevX.10.021041} {\bibfield  {journal} {\bibinfo  {journal} {Phys. Rev. X}\ }\textbf {\bibinfo {volume} {10}},\ \bibinfo {pages} {021041} (\bibinfo {year} {2020})}\BibitemShut {NoStop}%
\bibitem [{\citenamefont {Mil}\ \emph {et~al.}(2020)\citenamefont {Mil}, \citenamefont {Zache}, \citenamefont {Hegde}, \citenamefont {Xia}, \citenamefont {Bhatt}, \citenamefont {Oberthaler}, \citenamefont {Hauke}, \citenamefont {Berges},\ and\ \citenamefont {Jendrzejewski}}]{mil2020scalable}%
  \BibitemOpen
  \bibfield  {author} {\bibinfo {author} {\bibfnamefont {A.}~\bibnamefont {Mil}}, \bibinfo {author} {\bibfnamefont {T.~V.}\ \bibnamefont {Zache}}, \bibinfo {author} {\bibfnamefont {A.}~\bibnamefont {Hegde}}, \bibinfo {author} {\bibfnamefont {A.}~\bibnamefont {Xia}}, \bibinfo {author} {\bibfnamefont {R.~P.}\ \bibnamefont {Bhatt}}, \bibinfo {author} {\bibfnamefont {M.~K.}\ \bibnamefont {Oberthaler}}, \bibinfo {author} {\bibfnamefont {P.}~\bibnamefont {Hauke}}, \bibinfo {author} {\bibfnamefont {J.}~\bibnamefont {Berges}},\ and\ \bibinfo {author} {\bibfnamefont {F.}~\bibnamefont {Jendrzejewski}},\ }\bibfield  {title} {\bibinfo {title} {{A scalable realization of local U (1) gauge invariance in cold atomic mixtures}},\ }\href {https://doi.org/10.1126/science.aaz5312} {\bibfield  {journal} {\bibinfo  {journal} {Science}\ }\textbf {\bibinfo {volume} {367}},\ \bibinfo {pages} {1128} (\bibinfo {year} {2020})}\BibitemShut {NoStop}%
\bibitem [{\citenamefont {Yang}\ \emph {et~al.}(2020)\citenamefont {Yang}, \citenamefont {Sun}, \citenamefont {Ott}, \citenamefont {Wang}, \citenamefont {Zache}, \citenamefont {Halimeh}, \citenamefont {Yuan}, \citenamefont {Hauke},\ and\ \citenamefont {Pan}}]{yang2020observation}%
  \BibitemOpen
  \bibfield  {author} {\bibinfo {author} {\bibfnamefont {B.}~\bibnamefont {Yang}}, \bibinfo {author} {\bibfnamefont {H.}~\bibnamefont {Sun}}, \bibinfo {author} {\bibfnamefont {R.}~\bibnamefont {Ott}}, \bibinfo {author} {\bibfnamefont {H.-Y.}\ \bibnamefont {Wang}}, \bibinfo {author} {\bibfnamefont {T.~V.}\ \bibnamefont {Zache}}, \bibinfo {author} {\bibfnamefont {J.~C.}\ \bibnamefont {Halimeh}}, \bibinfo {author} {\bibfnamefont {Z.-S.}\ \bibnamefont {Yuan}}, \bibinfo {author} {\bibfnamefont {P.}~\bibnamefont {Hauke}},\ and\ \bibinfo {author} {\bibfnamefont {J.-W.}\ \bibnamefont {Pan}},\ }\bibfield  {title} {\bibinfo {title} {{Observation of gauge invariance in a 71-site Bose--Hubbard quantum simulator}},\ }\href {https://doi.org/10.1038/s41586-020-2910-8} {\bibfield  {journal} {\bibinfo  {journal} {Nature}\ }\textbf {\bibinfo {volume} {587}},\ \bibinfo {pages} {392} (\bibinfo {year} {2020})}\BibitemShut {NoStop}%
\bibitem [{\citenamefont {Zhou}\ \emph {et~al.}(2022)\citenamefont {Zhou}, \citenamefont {Su}, \citenamefont {Halimeh}, \citenamefont {Ott}, \citenamefont {Sun}, \citenamefont {Hauke}, \citenamefont {Yang}, \citenamefont {Yuan}, \citenamefont {Berges},\ and\ \citenamefont {Pan}}]{zhou2022thermalization}%
  \BibitemOpen
  \bibfield  {author} {\bibinfo {author} {\bibfnamefont {Z.-Y.}\ \bibnamefont {Zhou}}, \bibinfo {author} {\bibfnamefont {G.-X.}\ \bibnamefont {Su}}, \bibinfo {author} {\bibfnamefont {J.~C.}\ \bibnamefont {Halimeh}}, \bibinfo {author} {\bibfnamefont {R.}~\bibnamefont {Ott}}, \bibinfo {author} {\bibfnamefont {H.}~\bibnamefont {Sun}}, \bibinfo {author} {\bibfnamefont {P.}~\bibnamefont {Hauke}}, \bibinfo {author} {\bibfnamefont {B.}~\bibnamefont {Yang}}, \bibinfo {author} {\bibfnamefont {Z.-S.}\ \bibnamefont {Yuan}}, \bibinfo {author} {\bibfnamefont {J.}~\bibnamefont {Berges}},\ and\ \bibinfo {author} {\bibfnamefont {J.-W.}\ \bibnamefont {Pan}},\ }\bibfield  {title} {\bibinfo {title} {{Thermalization dynamics of a gauge theory on a quantum simulator}},\ }\href {https://doi.org/10.1126/science.abl6277} {\bibfield  {journal} {\bibinfo  {journal} {Science}\ }\textbf {\bibinfo {volume} {377}},\ \bibinfo {pages} {311} (\bibinfo {year} {2022})}\BibitemShut {NoStop}%
\bibitem [{\citenamefont {Bernien}\ \emph {et~al.}(2017)\citenamefont {Bernien}, \citenamefont {Schwartz}, \citenamefont {Keesling}, \citenamefont {Levine}, \citenamefont {Omran}, \citenamefont {Pichler}, \citenamefont {Choi}, \citenamefont {Zibrov}, \citenamefont {Endres}, \citenamefont {Greiner} \emph {et~al.}}]{bernien2017probing}%
  \BibitemOpen
  \bibfield  {author} {\bibinfo {author} {\bibfnamefont {H.}~\bibnamefont {Bernien}}, \bibinfo {author} {\bibfnamefont {S.}~\bibnamefont {Schwartz}}, \bibinfo {author} {\bibfnamefont {A.}~\bibnamefont {Keesling}}, \bibinfo {author} {\bibfnamefont {H.}~\bibnamefont {Levine}}, \bibinfo {author} {\bibfnamefont {A.}~\bibnamefont {Omran}}, \bibinfo {author} {\bibfnamefont {H.}~\bibnamefont {Pichler}}, \bibinfo {author} {\bibfnamefont {S.}~\bibnamefont {Choi}}, \bibinfo {author} {\bibfnamefont {A.~S.}\ \bibnamefont {Zibrov}}, \bibinfo {author} {\bibfnamefont {M.}~\bibnamefont {Endres}}, \bibinfo {author} {\bibfnamefont {M.}~\bibnamefont {Greiner}}, \emph {et~al.},\ }\bibfield  {title} {\bibinfo {title} {{Probing many-body dynamics on a 51-atom quantum simulator}},\ }\href {https://doi.org/10.1038/nature24622} {\bibfield  {journal} {\bibinfo  {journal} {Nature}\ }\textbf {\bibinfo {volume} {551}},\ \bibinfo {pages} {579} (\bibinfo {year} {2017})}\BibitemShut {NoStop}%
\bibitem [{\citenamefont {Byrnes}\ and\ \citenamefont {Yamamoto}(2006)}]{byrnes2006simulating}%
  \BibitemOpen
  \bibfield  {author} {\bibinfo {author} {\bibfnamefont {T.}~\bibnamefont {Byrnes}}\ and\ \bibinfo {author} {\bibfnamefont {Y.}~\bibnamefont {Yamamoto}},\ }\bibfield  {title} {\bibinfo {title} {Simulating lattice gauge theories on a quantum computer},\ }\href {https://doi.org/10.1103/PhysRevA.73.022328} {\bibfield  {journal} {\bibinfo  {journal} {Phys. Rev. A}\ }\textbf {\bibinfo {volume} {73}},\ \bibinfo {pages} {022328} (\bibinfo {year} {2006})}\BibitemShut {NoStop}%
\bibitem [{\citenamefont {Martinez}\ \emph {et~al.}(2016)\citenamefont {Martinez}, \citenamefont {Muschik}, \citenamefont {Schindler}, \citenamefont {Nigg}, \citenamefont {Erhard}, \citenamefont {Heyl}, \citenamefont {Hauke}, \citenamefont {Dalmonte}, \citenamefont {Monz}, \citenamefont {Zoller} \emph {et~al.}}]{martinez2016real}%
  \BibitemOpen
  \bibfield  {author} {\bibinfo {author} {\bibfnamefont {E.~A.}\ \bibnamefont {Martinez}}, \bibinfo {author} {\bibfnamefont {C.~A.}\ \bibnamefont {Muschik}}, \bibinfo {author} {\bibfnamefont {P.}~\bibnamefont {Schindler}}, \bibinfo {author} {\bibfnamefont {D.}~\bibnamefont {Nigg}}, \bibinfo {author} {\bibfnamefont {A.}~\bibnamefont {Erhard}}, \bibinfo {author} {\bibfnamefont {M.}~\bibnamefont {Heyl}}, \bibinfo {author} {\bibfnamefont {P.}~\bibnamefont {Hauke}}, \bibinfo {author} {\bibfnamefont {M.}~\bibnamefont {Dalmonte}}, \bibinfo {author} {\bibfnamefont {T.}~\bibnamefont {Monz}}, \bibinfo {author} {\bibfnamefont {P.}~\bibnamefont {Zoller}}, \emph {et~al.},\ }\bibfield  {title} {\bibinfo {title} {{Real-time dynamics of lattice gauge theories with a few-qubit quantum computer}},\ }\href {https://doi.org/10.1038/nature18318} {\bibfield  {journal} {\bibinfo  {journal} {Nature}\ }\textbf {\bibinfo {volume} {534}},\ \bibinfo {pages} {516} (\bibinfo {year} {2016})}\BibitemShut {NoStop}%
\bibitem [{\citenamefont {Nguyen}\ \emph {et~al.}(2022)\citenamefont {Nguyen}, \citenamefont {Tran}, \citenamefont {Zhu}, \citenamefont {Green}, \citenamefont {Alderete}, \citenamefont {Davoudi},\ and\ \citenamefont {Linke}}]{nguyen2022digital}%
  \BibitemOpen
  \bibfield  {author} {\bibinfo {author} {\bibfnamefont {N.~H.}\ \bibnamefont {Nguyen}}, \bibinfo {author} {\bibfnamefont {M.~C.}\ \bibnamefont {Tran}}, \bibinfo {author} {\bibfnamefont {Y.}~\bibnamefont {Zhu}}, \bibinfo {author} {\bibfnamefont {A.~M.}\ \bibnamefont {Green}}, \bibinfo {author} {\bibfnamefont {C.~H.}\ \bibnamefont {Alderete}}, \bibinfo {author} {\bibfnamefont {Z.}~\bibnamefont {Davoudi}},\ and\ \bibinfo {author} {\bibfnamefont {N.~M.}\ \bibnamefont {Linke}},\ }\bibfield  {title} {\bibinfo {title} {{Digital quantum simulation of the schwinger model and symmetry protection with trapped ions}},\ }\href {https://doi.org/10.1103/PRXQuantum.3.020324} {\bibfield  {journal} {\bibinfo  {journal} {PRX Quantum}\ }\textbf {\bibinfo {volume} {3}},\ \bibinfo {pages} {020324} (\bibinfo {year} {2022})}\BibitemShut {NoStop}%
\bibitem [{\citenamefont {Farrell}\ \emph {et~al.}(2024{\natexlab{a}})\citenamefont {Farrell}, \citenamefont {Illa}, \citenamefont {Ciavarella},\ and\ \citenamefont {Savage}}]{farrell2024scalable}%
  \BibitemOpen
  \bibfield  {author} {\bibinfo {author} {\bibfnamefont {R.~C.}\ \bibnamefont {Farrell}}, \bibinfo {author} {\bibfnamefont {M.}~\bibnamefont {Illa}}, \bibinfo {author} {\bibfnamefont {A.~N.}\ \bibnamefont {Ciavarella}},\ and\ \bibinfo {author} {\bibfnamefont {M.~J.}\ \bibnamefont {Savage}},\ }\bibfield  {title} {\bibinfo {title} {Scalable circuits for preparing ground states on digital quantum computers: The schwinger model vacuum on 100 qubits},\ }\href {https://doi.org/10.1103/PRXQuantum.5.020315} {\bibfield  {journal} {\bibinfo  {journal} {PRX Quantum}\ }\textbf {\bibinfo {volume} {5}},\ \bibinfo {pages} {020315} (\bibinfo {year} {2024}{\natexlab{a}})}\BibitemShut {NoStop}%
\bibitem [{\citenamefont {Farrell}\ \emph {et~al.}(2024{\natexlab{b}})\citenamefont {Farrell}, \citenamefont {Illa}, \citenamefont {Ciavarella},\ and\ \citenamefont {Savage}}]{farrell2024quantumsimulationshadrondynamics}%
  \BibitemOpen
  \bibfield  {author} {\bibinfo {author} {\bibfnamefont {R.~C.}\ \bibnamefont {Farrell}}, \bibinfo {author} {\bibfnamefont {M.}~\bibnamefont {Illa}}, \bibinfo {author} {\bibfnamefont {A.~N.}\ \bibnamefont {Ciavarella}},\ and\ \bibinfo {author} {\bibfnamefont {M.~J.}\ \bibnamefont {Savage}},\ }\bibfield  {title} {\bibinfo {title} {Quantum simulations of hadron dynamics in the schwinger model using 112 qubits},\ }\href {https://doi.org/10.1103/PhysRevD.109.114510} {\bibfield  {journal} {\bibinfo  {journal} {Phys. Rev. D}\ }\textbf {\bibinfo {volume} {109}},\ \bibinfo {pages} {114510} (\bibinfo {year} {2024}{\natexlab{b}})}\BibitemShut {NoStop}%
\bibitem [{\citenamefont {Nigg}\ \emph {et~al.}(2014)\citenamefont {Nigg}, \citenamefont {Mueller}, \citenamefont {Martinez}, \citenamefont {Schindler}, \citenamefont {Hennrich}, \citenamefont {Monz}, \citenamefont {Martin-Delgado},\ and\ \citenamefont {Blatt}}]{nigg2014quantum}%
  \BibitemOpen
  \bibfield  {author} {\bibinfo {author} {\bibfnamefont {D.}~\bibnamefont {Nigg}}, \bibinfo {author} {\bibfnamefont {M.}~\bibnamefont {Mueller}}, \bibinfo {author} {\bibfnamefont {E.~A.}\ \bibnamefont {Martinez}}, \bibinfo {author} {\bibfnamefont {P.}~\bibnamefont {Schindler}}, \bibinfo {author} {\bibfnamefont {M.}~\bibnamefont {Hennrich}}, \bibinfo {author} {\bibfnamefont {T.}~\bibnamefont {Monz}}, \bibinfo {author} {\bibfnamefont {M.~A.}\ \bibnamefont {Martin-Delgado}},\ and\ \bibinfo {author} {\bibfnamefont {R.}~\bibnamefont {Blatt}},\ }\bibfield  {title} {\bibinfo {title} {{Quantum computations on a topologically encoded qubit}},\ }\href {https://doi.org/10.1126/science.125374} {\bibfield  {journal} {\bibinfo  {journal} {Science}\ }\textbf {\bibinfo {volume} {345}},\ \bibinfo {pages} {302} (\bibinfo {year} {2014})}\BibitemShut {NoStop}%
\bibitem [{\citenamefont {Bluvstein}\ \emph {et~al.}(2024)\citenamefont {Bluvstein}, \citenamefont {Evered}, \citenamefont {Geim}, \citenamefont {Li}, \citenamefont {Zhou}, \citenamefont {Manovitz}, \citenamefont {Ebadi}, \citenamefont {Cain}, \citenamefont {Kalinowski}, \citenamefont {Hangleiter} \emph {et~al.}}]{bluvstein2024logical}%
  \BibitemOpen
  \bibfield  {author} {\bibinfo {author} {\bibfnamefont {D.}~\bibnamefont {Bluvstein}}, \bibinfo {author} {\bibfnamefont {S.~J.}\ \bibnamefont {Evered}}, \bibinfo {author} {\bibfnamefont {A.~A.}\ \bibnamefont {Geim}}, \bibinfo {author} {\bibfnamefont {S.~H.}\ \bibnamefont {Li}}, \bibinfo {author} {\bibfnamefont {H.}~\bibnamefont {Zhou}}, \bibinfo {author} {\bibfnamefont {T.}~\bibnamefont {Manovitz}}, \bibinfo {author} {\bibfnamefont {S.}~\bibnamefont {Ebadi}}, \bibinfo {author} {\bibfnamefont {M.}~\bibnamefont {Cain}}, \bibinfo {author} {\bibfnamefont {M.}~\bibnamefont {Kalinowski}}, \bibinfo {author} {\bibfnamefont {D.}~\bibnamefont {Hangleiter}}, \emph {et~al.},\ }\bibfield  {title} {\bibinfo {title} {{Logical quantum processor based on reconfigurable atom arrays}},\ }\href {https://doi.org/10.1038/s41586-023-06927-3} {\bibfield  {journal} {\bibinfo  {journal} {Nature}\ }\textbf {\bibinfo {volume} {626}},\ \bibinfo {pages} {58} (\bibinfo {year} {2024})}\BibitemShut {NoStop}%
\bibitem [{\citenamefont {Gupta}\ \emph {et~al.}(2024)\citenamefont {Gupta}, \citenamefont {Sundaresan}, \citenamefont {Alexander}, \citenamefont {Wood}, \citenamefont {Merkel}, \citenamefont {Healy}, \citenamefont {Hillenbrand}, \citenamefont {Jochym-O’Connor}, \citenamefont {Wootton}, \citenamefont {Yoder} \emph {et~al.}}]{gupta2024encoding}%
  \BibitemOpen
  \bibfield  {author} {\bibinfo {author} {\bibfnamefont {R.~S.}\ \bibnamefont {Gupta}}, \bibinfo {author} {\bibfnamefont {N.}~\bibnamefont {Sundaresan}}, \bibinfo {author} {\bibfnamefont {T.}~\bibnamefont {Alexander}}, \bibinfo {author} {\bibfnamefont {C.~J.}\ \bibnamefont {Wood}}, \bibinfo {author} {\bibfnamefont {S.~T.}\ \bibnamefont {Merkel}}, \bibinfo {author} {\bibfnamefont {M.~B.}\ \bibnamefont {Healy}}, \bibinfo {author} {\bibfnamefont {M.}~\bibnamefont {Hillenbrand}}, \bibinfo {author} {\bibfnamefont {T.}~\bibnamefont {Jochym-O’Connor}}, \bibinfo {author} {\bibfnamefont {J.~R.}\ \bibnamefont {Wootton}}, \bibinfo {author} {\bibfnamefont {T.~J.}\ \bibnamefont {Yoder}}, \emph {et~al.},\ }\bibfield  {title} {\bibinfo {title} {{Encoding a magic state with beyond break-even fidelity}},\ }\href {https://doi.org/10.1038/s41586-023-06846-3} {\bibfield  {journal} {\bibinfo  {journal} {Nature}\ }\textbf {\bibinfo {volume} {625}},\ \bibinfo {pages} {259} (\bibinfo {year} {2024})}\BibitemShut {NoStop}%
\bibitem [{\citenamefont {Chandrasekharan}\ and\ \citenamefont {Wiese}(1997)}]{chandrasekharan97}%
  \BibitemOpen
  \bibfield  {author} {\bibinfo {author} {\bibfnamefont {S.}~\bibnamefont {Chandrasekharan}}\ and\ \bibinfo {author} {\bibfnamefont {U.}~\bibnamefont {Wiese}},\ }\bibfield  {title} {\bibinfo {title} {{Quantum link models: A discrete approach to gauge theories}},\ }\href {https://doi.org/10.1016/s0550-3213(97)80041-7} {\bibfield  {journal} {\bibinfo  {journal} {Nuclear Physics B}\ }\textbf {\bibinfo {volume} {492}},\ \bibinfo {pages} {455} (\bibinfo {year} {1997})}\BibitemShut {NoStop}%
\bibitem [{\citenamefont {Aidelsburger}\ \emph {et~al.}(2021)\citenamefont {Aidelsburger}, \citenamefont {Barbiero}, \citenamefont {Bermudez}, \citenamefont {Chanda},\ and\ \citenamefont {Dauphin~{$et \, al.$}}}]{Aidelsburger2021}%
  \BibitemOpen
  \bibfield  {author} {\bibinfo {author} {\bibfnamefont {M.}~\bibnamefont {Aidelsburger}}, \bibinfo {author} {\bibfnamefont {L.}~\bibnamefont {Barbiero}}, \bibinfo {author} {\bibfnamefont {A.}~\bibnamefont {Bermudez}}, \bibinfo {author} {\bibfnamefont {T.}~\bibnamefont {Chanda}},\ and\ \bibinfo {author} {\bibfnamefont {A.}~\bibnamefont {Dauphin~{$et \, al.$}}},\ }\bibfield  {title} {\bibinfo {title} {Cold atoms meet lattice gauge theory},\ }\href {https://doi.org/10.1098/rsta.2021.0064} {\bibfield  {journal} {\bibinfo  {journal} {Phil. Tras. R. Soc. A}\ }\textbf {\bibinfo {volume} {380}},\ \bibinfo {pages} {20210064} (\bibinfo {year} {2021})}\BibitemShut {NoStop}%
\bibitem [{\citenamefont {Leone}\ \emph {et~al.}(2022)\citenamefont {Leone}, \citenamefont {Oliviero},\ and\ \citenamefont {Hamma}}]{Leone2022stabilizer}%
  \BibitemOpen
  \bibfield  {author} {\bibinfo {author} {\bibfnamefont {L.}~\bibnamefont {Leone}}, \bibinfo {author} {\bibfnamefont {S.~F.}\ \bibnamefont {Oliviero}},\ and\ \bibinfo {author} {\bibfnamefont {A.}~\bibnamefont {Hamma}},\ }\bibfield  {title} {\bibinfo {title} {Stabilizer r{\'e}nyi entropy},\ }\href {https://doi.org/10.1103/PhysRevLett.128.050402} {\bibfield  {journal} {\bibinfo  {journal} {Phys. Rev. Lett.}\ }\textbf {\bibinfo {volume} {128}},\ \bibinfo {pages} {050402} (\bibinfo {year} {2022})}\BibitemShut {NoStop}%
\bibitem [{\citenamefont {Ba{\~n}uls}(2023)}]{Banuls2023}%
  \BibitemOpen
  \bibfield  {author} {\bibinfo {author} {\bibfnamefont {M.~C.}\ \bibnamefont {Ba{\~n}uls}},\ }\bibfield  {title} {\bibinfo {title} {Tensor network algorithms: A route map},\ }\href {https://doi.org/10.1146/annurev-conmatphys-040721-022705} {\bibfield  {journal} {\bibinfo  {journal} {Annual Review of Condensed Matter Physics}\ }\textbf {\bibinfo {volume} {14}},\ \bibinfo {pages} {173} (\bibinfo {year} {2023})}\BibitemShut {NoStop}%
\bibitem [{\citenamefont {Ran}\ \emph {et~al.}(2020)\citenamefont {Ran}, \citenamefont {Tirrito}, \citenamefont {Peng}, \citenamefont {Chen}, \citenamefont {Tagliacozzo}, \citenamefont {Su},\ and\ \citenamefont {Lewenstein}}]{ran2020tensor}%
  \BibitemOpen
  \bibfield  {author} {\bibinfo {author} {\bibfnamefont {S.-J.}\ \bibnamefont {Ran}}, \bibinfo {author} {\bibfnamefont {E.}~\bibnamefont {Tirrito}}, \bibinfo {author} {\bibfnamefont {C.}~\bibnamefont {Peng}}, \bibinfo {author} {\bibfnamefont {X.}~\bibnamefont {Chen}}, \bibinfo {author} {\bibfnamefont {L.}~\bibnamefont {Tagliacozzo}}, \bibinfo {author} {\bibfnamefont {G.}~\bibnamefont {Su}},\ and\ \bibinfo {author} {\bibfnamefont {M.}~\bibnamefont {Lewenstein}},\ }\href {https://doi.org/10.1007/978-3-030-34489-4} {\emph {\bibinfo {title} {{Tensor network contractions: methods and applications to quantum many-body systems}}}}\ (\bibinfo  {publisher} {Springer Nature},\ \bibinfo {year} {2020})\BibitemShut {NoStop}%
\bibitem [{\citenamefont {Schollw{\"o}ck}(2011)}]{Schollwoeck2011}%
  \BibitemOpen
  \bibfield  {author} {\bibinfo {author} {\bibfnamefont {U.}~\bibnamefont {Schollw{\"o}ck}},\ }\bibfield  {title} {\bibinfo {title} {{The density-matrix renormalization group in the age of matrix product states}},\ }\href {https://doi.org/10.1016/j.aop.2010.09.012} {\bibfield  {journal} {\bibinfo  {journal} {Annals of physics}\ }\textbf {\bibinfo {volume} {326}},\ \bibinfo {pages} {96} (\bibinfo {year} {2011})}\BibitemShut {NoStop}%
\bibitem [{\citenamefont {Or{\'u}s}(2014)}]{Orus2014annphys}%
  \BibitemOpen
  \bibfield  {author} {\bibinfo {author} {\bibfnamefont {R.}~\bibnamefont {Or{\'u}s}},\ }\bibfield  {title} {\bibinfo {title} {{A practical introduction to tensor networks: Matrix product states and projected entangled pair states}},\ }\href {https://doi.org/10.1016/j.aop.2014.06.013} {\bibfield  {journal} {\bibinfo  {journal} {Annals of physics}\ }\textbf {\bibinfo {volume} {349}},\ \bibinfo {pages} {117} (\bibinfo {year} {2014})}\BibitemShut {NoStop}%
\bibitem [{\citenamefont {Haug}\ and\ \citenamefont {Piroli}(2023{\natexlab{a}})}]{haug2023quantifying}%
  \BibitemOpen
  \bibfield  {author} {\bibinfo {author} {\bibfnamefont {T.}~\bibnamefont {Haug}}\ and\ \bibinfo {author} {\bibfnamefont {L.}~\bibnamefont {Piroli}},\ }\bibfield  {title} {\bibinfo {title} {Quantifying nonstabilizerness of matrix product states},\ }\href {https://doi.org/10.1103/PhysRevB.107.035148} {\bibfield  {journal} {\bibinfo  {journal} {Phys. Rev. B}\ }\textbf {\bibinfo {volume} {107}},\ \bibinfo {pages} {035148} (\bibinfo {year} {2023}{\natexlab{a}})}\BibitemShut {NoStop}%
\bibitem [{\citenamefont {Tarabunga}\ \emph {et~al.}(2024)\citenamefont {Tarabunga}, \citenamefont {Tirrito}, \citenamefont {Bañuls},\ and\ \citenamefont {Dalmonte}}]{Tarabunga2024nonstabilizerness}%
  \BibitemOpen
  \bibfield  {author} {\bibinfo {author} {\bibfnamefont {P.~S.}\ \bibnamefont {Tarabunga}}, \bibinfo {author} {\bibfnamefont {E.}~\bibnamefont {Tirrito}}, \bibinfo {author} {\bibfnamefont {M.~C.}\ \bibnamefont {Bañuls}},\ and\ \bibinfo {author} {\bibfnamefont {M.}~\bibnamefont {Dalmonte}},\ }\bibfield  {title} {\bibinfo {title} {Nonstabilizerness via matrix product states in the pauli basis},\ }\href {https://doi.org/10.1103/PhysRevLett.133.010601} {\bibfield  {journal} {\bibinfo  {journal} {Phys. Rev. Lett.}\ }\textbf {\bibinfo {volume} {133}},\ \bibinfo {pages} {010601} (\bibinfo {year} {2024})}\BibitemShut {NoStop}%
\bibitem [{\citenamefont {Haug}\ and\ \citenamefont {Piroli}(2023{\natexlab{b}})}]{haug2023stabilizer}%
  \BibitemOpen
  \bibfield  {author} {\bibinfo {author} {\bibfnamefont {T.}~\bibnamefont {Haug}}\ and\ \bibinfo {author} {\bibfnamefont {L.}~\bibnamefont {Piroli}},\ }\bibfield  {title} {\bibinfo {title} {{Stabilizer entropies and nonstabilizerness monotones}},\ }\href {https://doi.org/10.22331/q-2023-08-28-1092} {\bibfield  {journal} {\bibinfo  {journal} {Quantum}\ }\textbf {\bibinfo {volume} {7}},\ \bibinfo {pages} {1092} (\bibinfo {year} {2023}{\natexlab{b}})}\BibitemShut {NoStop}%
\bibitem [{\citenamefont {Lami}\ and\ \citenamefont {Collura}(2023)}]{Lami2023}%
  \BibitemOpen
  \bibfield  {author} {\bibinfo {author} {\bibfnamefont {G.}~\bibnamefont {Lami}}\ and\ \bibinfo {author} {\bibfnamefont {M.}~\bibnamefont {Collura}},\ }\bibfield  {title} {\bibinfo {title} {{Nonstabilizerness via Perfect Pauli Sampling of Matrix Product States}},\ }\href {https://doi.org/10.1103/PhysRevLett.131.180401} {\bibfield  {journal} {\bibinfo  {journal} {Phys. Rev. Lett.}\ }\textbf {\bibinfo {volume} {131}},\ \bibinfo {pages} {180401} (\bibinfo {year} {2023})}\BibitemShut {NoStop}%
\bibitem [{\citenamefont {Tarabunga}(2024)}]{tarabunga2023critical}%
  \BibitemOpen
  \bibfield  {author} {\bibinfo {author} {\bibfnamefont {P.~S.}\ \bibnamefont {Tarabunga}},\ }\bibfield  {title} {\bibinfo {title} {Critical behaviors of non-stabilizerness in quantum spin chains},\ }\href {https://doi.org/10.22331/q-2024-07-17-1413} {\bibfield  {journal} {\bibinfo  {journal} {Quantum}\ }\textbf {\bibinfo {volume} {8}},\ \bibinfo {pages} {1413} (\bibinfo {year} {2024})}\BibitemShut {NoStop}%
\bibitem [{\citenamefont {Tarabunga}\ \emph {et~al.}(2023)\citenamefont {Tarabunga}, \citenamefont {Tirrito}, \citenamefont {Chanda},\ and\ \citenamefont {Dalmonte}}]{Tarabunga2023many}%
  \BibitemOpen
  \bibfield  {author} {\bibinfo {author} {\bibfnamefont {P.~S.}\ \bibnamefont {Tarabunga}}, \bibinfo {author} {\bibfnamefont {E.}~\bibnamefont {Tirrito}}, \bibinfo {author} {\bibfnamefont {T.}~\bibnamefont {Chanda}},\ and\ \bibinfo {author} {\bibfnamefont {M.}~\bibnamefont {Dalmonte}},\ }\bibfield  {title} {\bibinfo {title} {Many-body magic via pauli-markov chains—from criticality to gauge theories},\ }\href {https://doi.org/10.1103/PRXQuantum.4.040317} {\bibfield  {journal} {\bibinfo  {journal} {PRX Quantum}\ }\textbf {\bibinfo {volume} {4}},\ \bibinfo {pages} {040317} (\bibinfo {year} {2023})}\BibitemShut {NoStop}%
\bibitem [{\citenamefont {Ballarin}\ \emph {et~al.}(2024)\citenamefont {Ballarin}, \citenamefont {Silvi}, \citenamefont {Montangero},\ and\ \citenamefont {Jaschke}}]{ballarin2024optimalsamplingtensornetworks}%
  \BibitemOpen
  \bibfield  {author} {\bibinfo {author} {\bibfnamefont {M.}~\bibnamefont {Ballarin}}, \bibinfo {author} {\bibfnamefont {P.}~\bibnamefont {Silvi}}, \bibinfo {author} {\bibfnamefont {S.}~\bibnamefont {Montangero}},\ and\ \bibinfo {author} {\bibfnamefont {D.}~\bibnamefont {Jaschke}},\ }\href {https://arxiv.org/abs/2401.10330} {\bibinfo {title} {Optimal sampling of tensor networks targeting wave function's fast decaying tails}} (\bibinfo {year} {2024}),\ \Eprint {https://arxiv.org/abs/2401.10330} {arXiv:2401.10330 [quant-ph]} \BibitemShut {NoStop}%
\bibitem [{\citenamefont {Tarabunga}\ and\ \citenamefont {Castelnovo}(2024)}]{tarabunga2023magic}%
  \BibitemOpen
  \bibfield  {author} {\bibinfo {author} {\bibfnamefont {P.~S.}\ \bibnamefont {Tarabunga}}\ and\ \bibinfo {author} {\bibfnamefont {C.}~\bibnamefont {Castelnovo}},\ }\bibfield  {title} {\bibinfo {title} {Magic in generalized {Rokhsar-Kivelson} wavefunctions},\ }\href {https://doi.org/10.22331/q-2024-05-14-1347} {\bibfield  {journal} {\bibinfo  {journal} {Quantum}\ }\textbf {\bibinfo {volume} {8}},\ \bibinfo {pages} {1347} (\bibinfo {year} {2024})}\BibitemShut {NoStop}%
\bibitem [{\citenamefont {White}\ \emph {et~al.}(2021)\citenamefont {White}, \citenamefont {Cao},\ and\ \citenamefont {Swingle}}]{white2021conformal}%
  \BibitemOpen
  \bibfield  {author} {\bibinfo {author} {\bibfnamefont {C.~D.}\ \bibnamefont {White}}, \bibinfo {author} {\bibfnamefont {C.}~\bibnamefont {Cao}},\ and\ \bibinfo {author} {\bibfnamefont {B.}~\bibnamefont {Swingle}},\ }\bibfield  {title} {\bibinfo {title} {Conformal field theories are magical},\ }\href {https://doi.org/10.1103/PhysRevB.103.075145} {\bibfield  {journal} {\bibinfo  {journal} {Physical Review B}\ }\textbf {\bibinfo {volume} {103}},\ \bibinfo {pages} {075145} (\bibinfo {year} {2021})}\BibitemShut {NoStop}%
\bibitem [{\citenamefont {Schwinger}(1962)}]{Schwinger1}%
  \BibitemOpen
  \bibfield  {author} {\bibinfo {author} {\bibfnamefont {J.}~\bibnamefont {Schwinger}},\ }\bibfield  {title} {\bibinfo {title} {{Gauge invariance and mass. II}},\ }\href {https://doi.org/10.1103/PhysRev.128.2425} {\bibfield  {journal} {\bibinfo  {journal} {Phys. Rev.}\ }\textbf {\bibinfo {volume} {128}},\ \bibinfo {pages} {2425} (\bibinfo {year} {1962})}\BibitemShut {NoStop}%
\bibitem [{\citenamefont {Hamer}\ \emph {et~al.}(1997)\citenamefont {Hamer}, \citenamefont {Weihong},\ and\ \citenamefont {Oitmaa}}]{Encoding}%
  \BibitemOpen
  \bibfield  {author} {\bibinfo {author} {\bibfnamefont {C.~J.}\ \bibnamefont {Hamer}}, \bibinfo {author} {\bibfnamefont {Z.}~\bibnamefont {Weihong}},\ and\ \bibinfo {author} {\bibfnamefont {J.}~\bibnamefont {Oitmaa}},\ }\bibfield  {title} {\bibinfo {title} {{Series expansions for the massive Schwinger model in Hamiltonian lattice theory}},\ }\href {https://doi.org/10.1103/PhysRevD.56.55} {\bibfield  {journal} {\bibinfo  {journal} {Phys. Rev. D}\ }\textbf {\bibinfo {volume} {56}},\ \bibinfo {pages} {55} (\bibinfo {year} {1997})}\BibitemShut {NoStop}%
\bibitem [{\citenamefont {Horn}(1981)}]{QLink1}%
  \BibitemOpen
  \bibfield  {author} {\bibinfo {author} {\bibfnamefont {D.}~\bibnamefont {Horn}},\ }\bibfield  {title} {\bibinfo {title} {{Finite matrix models with continuous local gauge invariance}},\ }\href {https://doi.org/10.1016/0370-2693(81)90763-2} {\bibfield  {journal} {\bibinfo  {journal} {Physics Letters B}\ }\textbf {\bibinfo {volume} {100}},\ \bibinfo {pages} {149} (\bibinfo {year} {1981})}\BibitemShut {NoStop}%
\bibitem [{\citenamefont {Orland}\ and\ \citenamefont {Rohrlich}(1990)}]{QLink2}%
  \BibitemOpen
  \bibfield  {author} {\bibinfo {author} {\bibfnamefont {P.}~\bibnamefont {Orland}}\ and\ \bibinfo {author} {\bibfnamefont {D.}~\bibnamefont {Rohrlich}},\ }\bibfield  {title} {\bibinfo {title} {{Lattice gauge magnets: Local isospin from spin}},\ }\href {https://doi.org/10.1016/0550-3213(90)90646-U} {\bibfield  {journal} {\bibinfo  {journal} {Nucl. Phys. B}\ }\textbf {\bibinfo {volume} {338}},\ \bibinfo {pages} {647} (\bibinfo {year} {1990})}\BibitemShut {NoStop}%
\bibitem [{\citenamefont {Brower}\ \emph {et~al.}(1999)\citenamefont {Brower}, \citenamefont {Chandrasekharan},\ and\ \citenamefont {Wiese}}]{Brower1999}%
  \BibitemOpen
  \bibfield  {author} {\bibinfo {author} {\bibfnamefont {R.}~\bibnamefont {Brower}}, \bibinfo {author} {\bibfnamefont {S.}~\bibnamefont {Chandrasekharan}},\ and\ \bibinfo {author} {\bibfnamefont {U.}~\bibnamefont {Wiese}},\ }\bibfield  {title} {\bibinfo {title} {{QCD as a quantum link model}},\ }\href {https://doi.org/10.1103/PhysRevD.60.094502} {\bibfield  {journal} {\bibinfo  {journal} {Phys. Rev. D}\ }\textbf {\bibinfo {volume} {60}},\ \bibinfo {pages} {094502} (\bibinfo {year} {1999})}\BibitemShut {NoStop}%
\bibitem [{\citenamefont {Kogut}\ and\ \citenamefont {Susskind}(1975)}]{KogutSusskindFormulation}%
  \BibitemOpen
  \bibfield  {author} {\bibinfo {author} {\bibfnamefont {J.}~\bibnamefont {Kogut}}\ and\ \bibinfo {author} {\bibfnamefont {L.}~\bibnamefont {Susskind}},\ }\bibfield  {title} {\bibinfo {title} {{Hamiltonian formulation of {W}ilson's lattice gauge theories}},\ }\href {https://doi.org/10.1103/PhysRevD.11.395} {\bibfield  {journal} {\bibinfo  {journal} {Phys. Rev. D}\ }\textbf {\bibinfo {volume} {11}},\ \bibinfo {pages} {395} (\bibinfo {year} {1975})}\BibitemShut {NoStop}%
\bibitem [{\citenamefont {Fendley}\ \emph {et~al.}(2004)\citenamefont {Fendley}, \citenamefont {Sengupta},\ and\ \citenamefont {Sachdev}}]{fendley2004}%
  \BibitemOpen
  \bibfield  {author} {\bibinfo {author} {\bibfnamefont {P.}~\bibnamefont {Fendley}}, \bibinfo {author} {\bibfnamefont {K.}~\bibnamefont {Sengupta}},\ and\ \bibinfo {author} {\bibfnamefont {S.}~\bibnamefont {Sachdev}},\ }\bibfield  {title} {\bibinfo {title} {Competing density-wave orders in a one-dimensional hard-boson model},\ }\href {https://doi.org/10.1103/physrevb.69.075106} {\bibfield  {journal} {\bibinfo  {journal} {Phys. Rev. B}\ }\textbf {\bibinfo {volume} {69}},\ \bibinfo {pages} {075106} (\bibinfo {year} {2004})}\BibitemShut {NoStop}%
\bibitem [{\citenamefont {Mitsuhashi}\ and\ \citenamefont {Yoshioka}(2023)}]{mitsuhashi_clifford_2023}%
  \BibitemOpen
  \bibfield  {author} {\bibinfo {author} {\bibfnamefont {Y.}~\bibnamefont {Mitsuhashi}}\ and\ \bibinfo {author} {\bibfnamefont {N.}~\bibnamefont {Yoshioka}},\ }\bibfield  {title} {\bibinfo {title} {Clifford group and unitary designs under symmetry},\ }\href {https://doi.org/10.1103/PRXQuantum.4.040331} {\bibfield  {journal} {\bibinfo  {journal} {PRX Quantum}\ }\textbf {\bibinfo {volume} {4}},\ \bibinfo {pages} {040331} (\bibinfo {year} {2023})}\BibitemShut {NoStop}%
\bibitem [{\citenamefont {Leone}\ and\ \citenamefont {Bittel}(2024)}]{leone2024stabilizer}%
  \BibitemOpen
  \bibfield  {author} {\bibinfo {author} {\bibfnamefont {L.}~\bibnamefont {Leone}}\ and\ \bibinfo {author} {\bibfnamefont {L.}~\bibnamefont {Bittel}},\ }\href@noop {} {\bibinfo {title} {Stabilizer entropies are monotones for magic-state resource theory}} (\bibinfo {year} {2024}),\ \Eprint {https://arxiv.org/abs/2404.11652} {arXiv:2404.11652 [quant-ph]} \BibitemShut {NoStop}%
\bibitem [{\citenamefont {Passarelli}\ \emph {et~al.}(2024)\citenamefont {Passarelli}, \citenamefont {Fazio},\ and\ \citenamefont {Lucignano}}]{passarelli2024nonstabilizernesspermutationallyinvariantsystems}%
  \BibitemOpen
  \bibfield  {author} {\bibinfo {author} {\bibfnamefont {G.}~\bibnamefont {Passarelli}}, \bibinfo {author} {\bibfnamefont {R.}~\bibnamefont {Fazio}},\ and\ \bibinfo {author} {\bibfnamefont {P.}~\bibnamefont {Lucignano}},\ }\bibfield  {title} {\bibinfo {title} {Nonstabilizerness of permutationally invariant systems},\ }\href {https://doi.org/10.1103/PhysRevA.110.022436} {\bibfield  {journal} {\bibinfo  {journal} {Physical Review A}\ }\textbf {\bibinfo {volume} {110}},\ \bibinfo {pages} {022436} (\bibinfo {year} {2024})}\BibitemShut {NoStop}%
\bibitem [{\citenamefont {Liu}\ and\ \citenamefont {Clark}(2024)}]{liu2024nonequilibriumquantummontecarlo}%
  \BibitemOpen
  \bibfield  {author} {\bibinfo {author} {\bibfnamefont {Z.}~\bibnamefont {Liu}}\ and\ \bibinfo {author} {\bibfnamefont {B.~K.}\ \bibnamefont {Clark}},\ }\href@noop {} {\bibinfo {title} {Non-equilibrium quantum monte carlo algorithm for stabilizer r\'enyi entropy in spin systems}} (\bibinfo {year} {2024}),\ \Eprint {https://arxiv.org/abs/2405.19577} {arXiv:2405.19577 [quant-ph]} \BibitemShut {NoStop}%
\bibitem [{\citenamefont {Oliviero}\ \emph {et~al.}(2022{\natexlab{a}})\citenamefont {Oliviero}, \citenamefont {Leone},\ and\ \citenamefont {Hamma}}]{Oliviero2022}%
  \BibitemOpen
  \bibfield  {author} {\bibinfo {author} {\bibfnamefont {S.~F.~E.}\ \bibnamefont {Oliviero}}, \bibinfo {author} {\bibfnamefont {L.}~\bibnamefont {Leone}},\ and\ \bibinfo {author} {\bibfnamefont {A.}~\bibnamefont {Hamma}},\ }\bibfield  {title} {\bibinfo {title} {Magic-state resource theory for the ground state of the transverse-field ising model},\ }\href {https://doi.org/10.1103/physreva.106.042426} {\bibfield  {journal} {\bibinfo  {journal} {Phys. Rev.A}\ }\textbf {\bibinfo {volume} {106}},\ \bibinfo {pages} {042426} (\bibinfo {year} {2022}{\natexlab{a}})}\BibitemShut {NoStop}%
\bibitem [{\citenamefont {Frau}\ \emph {et~al.}(2024)\citenamefont {Frau}, \citenamefont {Tarabunga}, \citenamefont {Collura}, \citenamefont {Dalmonte},\ and\ \citenamefont {Tirrito}}]{Frau2024}%
  \BibitemOpen
  \bibfield  {author} {\bibinfo {author} {\bibfnamefont {M.}~\bibnamefont {Frau}}, \bibinfo {author} {\bibfnamefont {P.~S.}\ \bibnamefont {Tarabunga}}, \bibinfo {author} {\bibfnamefont {M.}~\bibnamefont {Collura}}, \bibinfo {author} {\bibfnamefont {M.}~\bibnamefont {Dalmonte}},\ and\ \bibinfo {author} {\bibfnamefont {E.}~\bibnamefont {Tirrito}},\ }\bibfield  {title} {\bibinfo {title} {Nonstabilizerness versus entanglement in matrix product states},\ }\href {https://doi.org/10.1103/physrevb.110.045101} {\bibfield  {journal} {\bibinfo  {journal} {Phys. Rev. B}\ }\textbf {\bibinfo {volume} {110}},\ \bibinfo {pages} {045101} (\bibinfo {year} {2024})}\BibitemShut {NoStop}%
\bibitem [{\citenamefont {Turkeshi}\ \emph {et~al.}(2024)\citenamefont {Turkeshi}, \citenamefont {Tirrito},\ and\ \citenamefont {Sierant}}]{turkeshi2024magicspreadingrandomquantum}%
  \BibitemOpen
  \bibfield  {author} {\bibinfo {author} {\bibfnamefont {X.}~\bibnamefont {Turkeshi}}, \bibinfo {author} {\bibfnamefont {E.}~\bibnamefont {Tirrito}},\ and\ \bibinfo {author} {\bibfnamefont {P.}~\bibnamefont {Sierant}},\ }\href@noop {} {\bibinfo {title} {Magic spreading in random quantum circuits}} (\bibinfo {year} {2024}),\ \Eprint {https://arxiv.org/abs/2407.03929} {arXiv:2407.03929 [quant-ph]} \BibitemShut {NoStop}%
\bibitem [{\citenamefont {Bejan}\ \emph {et~al.}(2024)\citenamefont {Bejan}, \citenamefont {McLauchlan},\ and\ \citenamefont {B{\'e}ri}}]{bejan2023dynamical}%
  \BibitemOpen
  \bibfield  {author} {\bibinfo {author} {\bibfnamefont {M.}~\bibnamefont {Bejan}}, \bibinfo {author} {\bibfnamefont {C.}~\bibnamefont {McLauchlan}},\ and\ \bibinfo {author} {\bibfnamefont {B.}~\bibnamefont {B{\'e}ri}},\ }\bibfield  {title} {\bibinfo {title} {Dynamical magic transitions in monitored clifford+ t circuits},\ }\href {https://doi.org/10.1103/PRXQuantum.5.030332} {\bibfield  {journal} {\bibinfo  {journal} {PRX Quantum}\ }\textbf {\bibinfo {volume} {5}},\ \bibinfo {pages} {030332} (\bibinfo {year} {2024})}\BibitemShut {NoStop}%
\bibitem [{\citenamefont {Fux}\ \emph {et~al.}(2023)\citenamefont {Fux}, \citenamefont {Tirrito}, \citenamefont {Dalmonte},\ and\ \citenamefont {Fazio}}]{fux2023entanglement}%
  \BibitemOpen
  \bibfield  {author} {\bibinfo {author} {\bibfnamefont {G.~E.}\ \bibnamefont {Fux}}, \bibinfo {author} {\bibfnamefont {E.}~\bibnamefont {Tirrito}}, \bibinfo {author} {\bibfnamefont {M.}~\bibnamefont {Dalmonte}},\ and\ \bibinfo {author} {\bibfnamefont {R.}~\bibnamefont {Fazio}},\ }\href@noop {} {\bibinfo {title} {Entanglement-magic separation in hybrid quantum circuits}} (\bibinfo {year} {2023}),\ \Eprint {https://arxiv.org/abs/2312.02039} {arXiv:2312.02039 [quant-ph]} \BibitemShut {NoStop}%
\bibitem [{\citenamefont {Tarabunga}\ and\ \citenamefont {Tirrito}(2024)}]{tarabunga2024magictransitionmeasurementonlycircuits}%
  \BibitemOpen
  \bibfield  {author} {\bibinfo {author} {\bibfnamefont {P.~S.}\ \bibnamefont {Tarabunga}}\ and\ \bibinfo {author} {\bibfnamefont {E.}~\bibnamefont {Tirrito}},\ }\href@noop {} {\bibinfo {title} {Magic transition in measurement-only circuits}} (\bibinfo {year} {2024}),\ \Eprint {https://arxiv.org/abs/2407.15939} {arXiv:2407.15939 [quant-ph]} \BibitemShut {NoStop}%
\bibitem [{\citenamefont {López}\ and\ \citenamefont {Kos}(2024)}]{lópez2024exact}%
  \BibitemOpen
  \bibfield  {author} {\bibinfo {author} {\bibfnamefont {J.~A.~M.}\ \bibnamefont {López}}\ and\ \bibinfo {author} {\bibfnamefont {P.}~\bibnamefont {Kos}},\ }\href@noop {} {\bibinfo {title} {Exact solution of long-range stabilizer r\'enyi entropy in the dual-unitary xxz model}} (\bibinfo {year} {2024}),\ \Eprint {https://arxiv.org/abs/2405.04448} {arXiv:2405.04448 [quant-ph]} \BibitemShut {NoStop}%
\bibitem [{\citenamefont {Paviglianiti}\ \emph {et~al.}(2024)\citenamefont {Paviglianiti}, \citenamefont {Lami}, \citenamefont {Collura},\ and\ \citenamefont {Silva}}]{paviglianiti2024estimatingnonstabilizernessdynamicssimulating}%
  \BibitemOpen
  \bibfield  {author} {\bibinfo {author} {\bibfnamefont {A.}~\bibnamefont {Paviglianiti}}, \bibinfo {author} {\bibfnamefont {G.}~\bibnamefont {Lami}}, \bibinfo {author} {\bibfnamefont {M.}~\bibnamefont {Collura}},\ and\ \bibinfo {author} {\bibfnamefont {A.}~\bibnamefont {Silva}},\ }\href@noop {} {\bibinfo {title} {Estimating non-stabilizerness dynamics without simulating it}} (\bibinfo {year} {2024}),\ \Eprint {https://arxiv.org/abs/2405.06054} {arXiv:2405.06054 [quant-ph]} \BibitemShut {NoStop}%
\bibitem [{\citenamefont {Lami}\ and\ \citenamefont {Collura}(2024)}]{lami2024learningstabilizergroupmatrix}%
  \BibitemOpen
  \bibfield  {author} {\bibinfo {author} {\bibfnamefont {G.}~\bibnamefont {Lami}}\ and\ \bibinfo {author} {\bibfnamefont {M.}~\bibnamefont {Collura}},\ }\href@noop {} {\bibinfo {title} {Learning the stabilizer group of a matrix product state}} (\bibinfo {year} {2024}),\ \Eprint {https://arxiv.org/abs/2401.16481} {arXiv:2401.16481 [quant-ph]} \BibitemShut {NoStop}%
\bibitem [{\citenamefont {Mello}\ \emph {et~al.}(2024)\citenamefont {Mello}, \citenamefont {Santini}, \citenamefont {Lami}, \citenamefont {De~Nardis},\ and\ \citenamefont {Collura}}]{mello2024clifford}%
  \BibitemOpen
  \bibfield  {author} {\bibinfo {author} {\bibfnamefont {A.~F.}\ \bibnamefont {Mello}}, \bibinfo {author} {\bibfnamefont {A.}~\bibnamefont {Santini}}, \bibinfo {author} {\bibfnamefont {G.}~\bibnamefont {Lami}}, \bibinfo {author} {\bibfnamefont {J.}~\bibnamefont {De~Nardis}},\ and\ \bibinfo {author} {\bibfnamefont {M.}~\bibnamefont {Collura}},\ }\href@noop {} {\bibinfo {title} {Clifford dressed time-dependent variational principle}} (\bibinfo {year} {2024}),\ \Eprint {https://arxiv.org/abs/2407.01692} {arXiv:2407.01692 [quant-ph]} \BibitemShut {NoStop}%
\bibitem [{\citenamefont {Ahmadi}\ \emph {et~al.}(2024)\citenamefont {Ahmadi}, \citenamefont {Helsen}, \citenamefont {Karaca},\ and\ \citenamefont {Greplova}}]{ahmadi2024mutual}%
  \BibitemOpen
  \bibfield  {author} {\bibinfo {author} {\bibfnamefont {A.}~\bibnamefont {Ahmadi}}, \bibinfo {author} {\bibfnamefont {J.}~\bibnamefont {Helsen}}, \bibinfo {author} {\bibfnamefont {C.}~\bibnamefont {Karaca}},\ and\ \bibinfo {author} {\bibfnamefont {E.}~\bibnamefont {Greplova}},\ }\href@noop {} {\bibinfo {title} {Mutual information fluctuations and non-stabilizerness in random circuits}} (\bibinfo {year} {2024}),\ \Eprint {https://arxiv.org/abs/2408.03831} {arXiv:2408.03831 [hep-lat]} \BibitemShut {NoStop}%
\bibitem [{\citenamefont {Oliviero}\ \emph {et~al.}(2022{\natexlab{b}})\citenamefont {Oliviero}, \citenamefont {Leone}, \citenamefont {Hamma},\ and\ \citenamefont {Lloyd}}]{Oliviero_2022}%
  \BibitemOpen
  \bibfield  {author} {\bibinfo {author} {\bibfnamefont {S.~F.~E.}\ \bibnamefont {Oliviero}}, \bibinfo {author} {\bibfnamefont {L.}~\bibnamefont {Leone}}, \bibinfo {author} {\bibfnamefont {A.}~\bibnamefont {Hamma}},\ and\ \bibinfo {author} {\bibfnamefont {S.}~\bibnamefont {Lloyd}},\ }\bibfield  {title} {\bibinfo {title} {Measuring magic on a quantum processor},\ }\href {https://doi.org/10.1038/s41534-022-00666-5} {\bibfield  {journal} {\bibinfo  {journal} {npj Quantum Information}\ }\textbf {\bibinfo {volume} {8}},\ \bibinfo {pages} {148} (\bibinfo {year} {2022}{\natexlab{b}})}\BibitemShut {NoStop}%
\bibitem [{\citenamefont {Haug}\ and\ \citenamefont {Kim}(2023)}]{haug2023scalable}%
  \BibitemOpen
  \bibfield  {author} {\bibinfo {author} {\bibfnamefont {T.}~\bibnamefont {Haug}}\ and\ \bibinfo {author} {\bibfnamefont {M.}~\bibnamefont {Kim}},\ }\bibfield  {title} {\bibinfo {title} {Scalable measures of magic resource for quantum computers},\ }\href {https://doi.org/10.1103/PRXQuantum.4.010301} {\bibfield  {journal} {\bibinfo  {journal} {PRX Quantum}\ }\textbf {\bibinfo {volume} {4}},\ \bibinfo {pages} {010301} (\bibinfo {year} {2023})}\BibitemShut {NoStop}%
\bibitem [{\citenamefont {Tirrito}\ \emph {et~al.}(2024)\citenamefont {Tirrito}, \citenamefont {Tarabunga}, \citenamefont {Lami}, \citenamefont {Chanda}, \citenamefont {Leone}, \citenamefont {Oliviero}, \citenamefont {Dalmonte}, \citenamefont {Collura},\ and\ \citenamefont {Hamma}}]{tirrito2023}%
  \BibitemOpen
  \bibfield  {author} {\bibinfo {author} {\bibfnamefont {E.}~\bibnamefont {Tirrito}}, \bibinfo {author} {\bibfnamefont {P.~S.}\ \bibnamefont {Tarabunga}}, \bibinfo {author} {\bibfnamefont {G.}~\bibnamefont {Lami}}, \bibinfo {author} {\bibfnamefont {T.}~\bibnamefont {Chanda}}, \bibinfo {author} {\bibfnamefont {L.}~\bibnamefont {Leone}}, \bibinfo {author} {\bibfnamefont {S.~F.}\ \bibnamefont {Oliviero}}, \bibinfo {author} {\bibfnamefont {M.}~\bibnamefont {Dalmonte}}, \bibinfo {author} {\bibfnamefont {M.}~\bibnamefont {Collura}},\ and\ \bibinfo {author} {\bibfnamefont {A.}~\bibnamefont {Hamma}},\ }\bibfield  {title} {\bibinfo {title} {Quantifying nonstabilizerness through entanglement spectrum flatness},\ }\href {https://doi.org/10.1103/physreva.109.l040401} {\bibfield  {journal} {\bibinfo  {journal} {Phys. Rev. A}\ }\textbf {\bibinfo {volume} {109}},\ \bibinfo {pages} {L040401} (\bibinfo {year} {2024})}\BibitemShut {NoStop}%
\bibitem [{\citenamefont {Haug}\ \emph {et~al.}(2024)\citenamefont {Haug}, \citenamefont {Lee},\ and\ \citenamefont {Kim}}]{Haug2024}%
  \BibitemOpen
  \bibfield  {author} {\bibinfo {author} {\bibfnamefont {T.}~\bibnamefont {Haug}}, \bibinfo {author} {\bibfnamefont {S.}~\bibnamefont {Lee}},\ and\ \bibinfo {author} {\bibfnamefont {M.}~\bibnamefont {Kim}},\ }\bibfield  {title} {\bibinfo {title} {Efficient quantum algorithms for stabilizer entropies},\ }\href {https://doi.org/10.1103/PhysRevLett.132.240602} {\bibfield  {journal} {\bibinfo  {journal} {Phys. Rev. Lett.}\ }\textbf {\bibinfo {volume} {132}},\ \bibinfo {pages} {240602} (\bibinfo {year} {2024})}\BibitemShut {NoStop}%
\bibitem [{\citenamefont {Niroula}\ \emph {et~al.}(2023)\citenamefont {Niroula}, \citenamefont {White}, \citenamefont {Wang}, \citenamefont {Johri}, \citenamefont {Zhu}, \citenamefont {Monroe}, \citenamefont {Noel},\ and\ \citenamefont {Gullans}}]{niroula2023phase}%
  \BibitemOpen
  \bibfield  {author} {\bibinfo {author} {\bibfnamefont {P.}~\bibnamefont {Niroula}}, \bibinfo {author} {\bibfnamefont {C.~D.}\ \bibnamefont {White}}, \bibinfo {author} {\bibfnamefont {Q.}~\bibnamefont {Wang}}, \bibinfo {author} {\bibfnamefont {S.}~\bibnamefont {Johri}}, \bibinfo {author} {\bibfnamefont {D.}~\bibnamefont {Zhu}}, \bibinfo {author} {\bibfnamefont {C.}~\bibnamefont {Monroe}}, \bibinfo {author} {\bibfnamefont {C.}~\bibnamefont {Noel}},\ and\ \bibinfo {author} {\bibfnamefont {M.~J.}\ \bibnamefont {Gullans}},\ }\href@noop {} {\bibinfo {title} {Phase transition in magic with random quantum circuits}} (\bibinfo {year} {2023}),\ \Eprint {https://arxiv.org/abs/2304.10481} {arXiv:2304.10481 [quant-ph]} \BibitemShut {NoStop}%
\bibitem [{\citenamefont {Ardonne}\ \emph {et~al.}(2004)\citenamefont {Ardonne}, \citenamefont {Fendley},\ and\ \citenamefont {Fradkin}}]{Ardonne2004}%
  \BibitemOpen
  \bibfield  {author} {\bibinfo {author} {\bibfnamefont {E.}~\bibnamefont {Ardonne}}, \bibinfo {author} {\bibfnamefont {P.}~\bibnamefont {Fendley}},\ and\ \bibinfo {author} {\bibfnamefont {E.}~\bibnamefont {Fradkin}},\ }\bibfield  {title} {\bibinfo {title} {Topological order and conformal quantum critical points},\ }\href {https://doi.org/10.1016/j.aop.2004.01.004} {\bibfield  {journal} {\bibinfo  {journal} {Annals of Physics}\ }\textbf {\bibinfo {volume} {310}},\ \bibinfo {pages} {493} (\bibinfo {year} {2004})}\BibitemShut {NoStop}%
\bibitem [{\citenamefont {Henley}(2004)}]{Henley2004}%
  \BibitemOpen
  \bibfield  {author} {\bibinfo {author} {\bibfnamefont {C.~L.}\ \bibnamefont {Henley}},\ }\bibfield  {title} {\bibinfo {title} {{From classical to quantum dynamics at Rokhsar{\textendash}Kivelson points}},\ }\href {https://doi.org/10.1088/0953-8984/16/11/045} {\bibfield  {journal} {\bibinfo  {journal} {Journal of Physics: Condensed Matter}\ }\textbf {\bibinfo {volume} {16}},\ \bibinfo {pages} {S891} (\bibinfo {year} {2004})}\BibitemShut {NoStop}%
\bibitem [{\citenamefont {Castelnovo}\ \emph {et~al.}(2005)\citenamefont {Castelnovo}, \citenamefont {Chamon}, \citenamefont {Mudry},\ and\ \citenamefont {Pujol}}]{Castelnovo2005}%
  \BibitemOpen
  \bibfield  {author} {\bibinfo {author} {\bibfnamefont {C.}~\bibnamefont {Castelnovo}}, \bibinfo {author} {\bibfnamefont {C.}~\bibnamefont {Chamon}}, \bibinfo {author} {\bibfnamefont {C.}~\bibnamefont {Mudry}},\ and\ \bibinfo {author} {\bibfnamefont {P.}~\bibnamefont {Pujol}},\ }\bibfield  {title} {\bibinfo {title} {{From quantum mechanics to classical statistical physics: Generalized Rokhsar{\textendash}Kivelson Hamiltonians and the {\textquotedblleft}Stochastic Matrix Form{\textquotedblright} decomposition}},\ }\href {https://doi.org/10.1016/j.aop.2005.01.006} {\bibfield  {journal} {\bibinfo  {journal} {Annals of Physics}\ }\textbf {\bibinfo {volume} {318}},\ \bibinfo {pages} {316} (\bibinfo {year} {2005})}\BibitemShut {NoStop}%
\bibitem [{\citenamefont {Okunishi}\ \emph {et~al.}(2022)\citenamefont {Okunishi}, \citenamefont {Nishino},\ and\ \citenamefont {Ueda}}]{Okunishi2022}%
  \BibitemOpen
  \bibfield  {author} {\bibinfo {author} {\bibfnamefont {K.}~\bibnamefont {Okunishi}}, \bibinfo {author} {\bibfnamefont {T.}~\bibnamefont {Nishino}},\ and\ \bibinfo {author} {\bibfnamefont {H.}~\bibnamefont {Ueda}},\ }\bibfield  {title} {\bibinfo {title} {Developments in the tensor network — from statistical mechanics to quantum entanglement},\ }\href {https://doi.org/10.7566/jpsj.91.062001} {\bibfield  {journal} {\bibinfo  {journal} {Journal of the Physical Society of Japan}\ }\textbf {\bibinfo {volume} {91}},\ \bibinfo {pages} {062001} (\bibinfo {year} {2022})}\BibitemShut {NoStop}%
\bibitem [{sup()}]{supmat}%
  \BibitemOpen
  \href@noop {} {\bibinfo  {journal} {See Supplementary Material at [URL will be inserted by publisher] for further details of our analysis.}\ }\BibitemShut {NoStop}%
\bibitem [{\citenamefont {{\L}{\c{a}}cki}\ \emph {et~al.}(2016)\citenamefont {{\L}{\c{a}}cki}, \citenamefont {Damski},\ and\ \citenamefont {Zakrzewski}}]{Lacki2016}%
  \BibitemOpen
\bibfield  {journal} {  }\bibfield  {author} {\bibinfo {author} {\bibfnamefont {M.}~\bibnamefont {{\L}{\c{a}}cki}}, \bibinfo {author} {\bibfnamefont {B.}~\bibnamefont {Damski}},\ and\ \bibinfo {author} {\bibfnamefont {J.}~\bibnamefont {Zakrzewski}},\ }\bibfield  {title} {\bibinfo {title} {{Locating the quantum critical point of the Bose-Hubbard model through singularities of simple observables}},\ }\href {https://doi.org/10.1038/srep38340} {\bibfield  {journal} {\bibinfo  {journal} {Scientific Reports}\ }\textbf {\bibinfo {volume} {6}},\ \bibinfo {pages} {38340} (\bibinfo {year} {2016})}\BibitemShut {NoStop}%
\bibitem [{\citenamefont {\L{}\c{a}cki}\ and\ \citenamefont {Damski}(2021)}]{Lacki2021}%
  \BibitemOpen
  \bibfield  {author} {\bibinfo {author} {\bibfnamefont {M.}~\bibnamefont {\L{}\c{a}cki}}\ and\ \bibinfo {author} {\bibfnamefont {B.}~\bibnamefont {Damski}},\ }\bibfield  {title} {\bibinfo {title} {{Evidence from on-site atom number fluctuations for a quantum Berezinskii-Kosterlitz-Thouless transition in the one-dimensional Bose-Hubbard model}},\ }\href {https://doi.org/10.1103/PhysRevB.104.155113} {\bibfield  {journal} {\bibinfo  {journal} {Phys. Rev. B}\ }\textbf {\bibinfo {volume} {104}},\ \bibinfo {pages} {155113} (\bibinfo {year} {2021})}\BibitemShut {NoStop}%
\bibitem [{Note1()}]{Note1}%
  \BibitemOpen
  \bibinfo {note} {The sign of the derivative depends on which direction the transition is approached and is thus non-generic.}\BibitemShut {Stop}%
\bibitem [{\citenamefont {Lamm}\ \emph {et~al.}(2019)\citenamefont {Lamm}, \citenamefont {Lawrence},\ and\ \citenamefont {Yamauchi}}]{Lamm19}%
  \BibitemOpen
  \bibfield  {author} {\bibinfo {author} {\bibfnamefont {H.}~\bibnamefont {Lamm}}, \bibinfo {author} {\bibfnamefont {S.}~\bibnamefont {Lawrence}},\ and\ \bibinfo {author} {\bibfnamefont {Y.}~\bibnamefont {Yamauchi}} (\bibinfo {collaboration} {NuQS Collaboration}),\ }\bibfield  {title} {\bibinfo {title} {{General methods for digital quantum simulation of gauge theories}},\ }\href {https://doi.org/10.1103/PhysRevD.100.034518} {\bibfield  {journal} {\bibinfo  {journal} {Phys. Rev. D}\ }\textbf {\bibinfo {volume} {100}},\ \bibinfo {pages} {034518} (\bibinfo {year} {2019})}\BibitemShut {NoStop}%
\bibitem [{\citenamefont {Murairi}\ \emph {et~al.}(2022)\citenamefont {Murairi}, \citenamefont {Cervia}, \citenamefont {Kumar}, \citenamefont {Bedaque},\ and\ \citenamefont {Alexandru}}]{Murairi22}%
  \BibitemOpen
  \bibfield  {author} {\bibinfo {author} {\bibfnamefont {E.~M.}\ \bibnamefont {Murairi}}, \bibinfo {author} {\bibfnamefont {M.~J.}\ \bibnamefont {Cervia}}, \bibinfo {author} {\bibfnamefont {H.}~\bibnamefont {Kumar}}, \bibinfo {author} {\bibfnamefont {P.~F.}\ \bibnamefont {Bedaque}},\ and\ \bibinfo {author} {\bibfnamefont {A.}~\bibnamefont {Alexandru}},\ }\bibfield  {title} {\bibinfo {title} {{How many quantum gates do gauge theories require?}},\ }\href {https://doi.org/10.1103/PhysRevD.106.094504} {\bibfield  {journal} {\bibinfo  {journal} {Phys. Rev. D}\ }\textbf {\bibinfo {volume} {106}},\ \bibinfo {pages} {094504} (\bibinfo {year} {2022})}\BibitemShut {NoStop}%
\bibitem [{\citenamefont {Ciavarella}\ and\ \citenamefont {Chernyshev}(2022)}]{ciavarella2022preparation}%
  \BibitemOpen
  \bibfield  {author} {\bibinfo {author} {\bibfnamefont {A.~N.}\ \bibnamefont {Ciavarella}}\ and\ \bibinfo {author} {\bibfnamefont {I.~A.}\ \bibnamefont {Chernyshev}},\ }\bibfield  {title} {\bibinfo {title} {{Preparation of the SU (3) lattice Yang-Mills vacuum with variational quantum methods}},\ }\href {https://doi.org/10.1103/PhysRevD.105.074504} {\bibfield  {journal} {\bibinfo  {journal} {Phys. Rev. D}\ }\textbf {\bibinfo {volume} {105}},\ \bibinfo {pages} {074504} (\bibinfo {year} {2022})}\BibitemShut {NoStop}%
\bibitem [{\citenamefont {Davoudi}\ \emph {et~al.}(2023)\citenamefont {Davoudi}, \citenamefont {Shaw},\ and\ \citenamefont {Stryker}}]{Davoudi23}%
  \BibitemOpen
  \bibfield  {author} {\bibinfo {author} {\bibfnamefont {Z.}~\bibnamefont {Davoudi}}, \bibinfo {author} {\bibfnamefont {A.~F.}\ \bibnamefont {Shaw}},\ and\ \bibinfo {author} {\bibfnamefont {J.~R.}\ \bibnamefont {Stryker}},\ }\bibfield  {title} {\bibinfo {title} {{General quantum algorithms for Hamiltonian simulation with applications to a non-Abelian lattice gauge theory}},\ }\href {https://doi.org/10.22331/q-2023-12-20-1213} {\bibfield  {journal} {\bibinfo  {journal} {Quantum}\ }\textbf {\bibinfo {volume} {7}},\ \bibinfo {pages} {1213} (\bibinfo {year} {2023})}\BibitemShut {NoStop}%
\bibitem [{\citenamefont {Osterloh}\ \emph {et~al.}(2005)\citenamefont {Osterloh}, \citenamefont {Baig}, \citenamefont {Santos}, \citenamefont {Zoller},\ and\ \citenamefont {Lewenstein}}]{osterloh2005cold}%
  \BibitemOpen
  \bibfield  {author} {\bibinfo {author} {\bibfnamefont {K.}~\bibnamefont {Osterloh}}, \bibinfo {author} {\bibfnamefont {M.}~\bibnamefont {Baig}}, \bibinfo {author} {\bibfnamefont {L.}~\bibnamefont {Santos}}, \bibinfo {author} {\bibfnamefont {P.}~\bibnamefont {Zoller}},\ and\ \bibinfo {author} {\bibfnamefont {M.}~\bibnamefont {Lewenstein}},\ }\bibfield  {title} {\bibinfo {title} {Cold atoms in non-abelian gauge potentials: From the hofstadter "moth" to lattice gauge theory},\ }\href {https://doi.org/10.1103/PhysRevLett.95.010403} {\bibfield  {journal} {\bibinfo  {journal} {Phys. Rev. Lett.}\ }\textbf {\bibinfo {volume} {95}},\ \bibinfo {pages} {010403} (\bibinfo {year} {2005})}\BibitemShut {NoStop}%
\bibitem [{\citenamefont {Banerjee}\ \emph {et~al.}(2013)\citenamefont {Banerjee}, \citenamefont {B{\"o}gli}, \citenamefont {Dalmonte}, \citenamefont {Rico}, \citenamefont {Stebler}, \citenamefont {Wiese},\ and\ \citenamefont {Zoller}}]{banerjee2013atomic}%
  \BibitemOpen
  \bibfield  {author} {\bibinfo {author} {\bibfnamefont {D.}~\bibnamefont {Banerjee}}, \bibinfo {author} {\bibfnamefont {M.}~\bibnamefont {B{\"o}gli}}, \bibinfo {author} {\bibfnamefont {M.}~\bibnamefont {Dalmonte}}, \bibinfo {author} {\bibfnamefont {E.}~\bibnamefont {Rico}}, \bibinfo {author} {\bibfnamefont {P.}~\bibnamefont {Stebler}}, \bibinfo {author} {\bibfnamefont {U.-J.}\ \bibnamefont {Wiese}},\ and\ \bibinfo {author} {\bibfnamefont {P.}~\bibnamefont {Zoller}},\ }\bibfield  {title} {\bibinfo {title} {Atomic quantum simulation of u(n) and su(n) non-abelian lattice gauge theories},\ }\href {https://doi.org/10.1103/PhysRevLett.110.125303} {\bibfield  {journal} {\bibinfo  {journal} {Phys. Rev. Lett.}\ }\textbf {\bibinfo {volume} {110}},\ \bibinfo {pages} {125303} (\bibinfo {year} {2013})}\BibitemShut {NoStop}%
\bibitem [{\citenamefont {Tagliacozzo}\ \emph {et~al.}(2013)\citenamefont {Tagliacozzo}, \citenamefont {Celi}, \citenamefont {Orland}, \citenamefont {Mitchell},\ and\ \citenamefont {Lewenstein}}]{tagliacozzo2013simulation}%
  \BibitemOpen
  \bibfield  {author} {\bibinfo {author} {\bibfnamefont {L.}~\bibnamefont {Tagliacozzo}}, \bibinfo {author} {\bibfnamefont {A.}~\bibnamefont {Celi}}, \bibinfo {author} {\bibfnamefont {P.}~\bibnamefont {Orland}}, \bibinfo {author} {\bibfnamefont {M.}~\bibnamefont {Mitchell}},\ and\ \bibinfo {author} {\bibfnamefont {M.}~\bibnamefont {Lewenstein}},\ }\bibfield  {title} {\bibinfo {title} {Simulation of non-abelian gauge theories with optical lattices},\ }\href {https://doi.org/10.1038/ncomms3615} {\bibfield  {journal} {\bibinfo  {journal} {Nature communications}\ }\textbf {\bibinfo {volume} {4}},\ \bibinfo {pages} {2615} (\bibinfo {year} {2013})}\BibitemShut {NoStop}%
\bibitem [{\citenamefont {Mezzacapo}\ \emph {et~al.}(2015)\citenamefont {Mezzacapo}, \citenamefont {Rico}, \citenamefont {Sab{\'\i}n}, \citenamefont {Egusquiza}, \citenamefont {Lamata},\ and\ \citenamefont {Solano}}]{mezzacapo2015non}%
  \BibitemOpen
  \bibfield  {author} {\bibinfo {author} {\bibfnamefont {A.}~\bibnamefont {Mezzacapo}}, \bibinfo {author} {\bibfnamefont {E.}~\bibnamefont {Rico}}, \bibinfo {author} {\bibfnamefont {C.}~\bibnamefont {Sab{\'\i}n}}, \bibinfo {author} {\bibfnamefont {I.}~\bibnamefont {Egusquiza}}, \bibinfo {author} {\bibfnamefont {L.}~\bibnamefont {Lamata}},\ and\ \bibinfo {author} {\bibfnamefont {E.}~\bibnamefont {Solano}},\ }\bibfield  {title} {\bibinfo {title} {Non-abelian su(2) lattice gauge theories in superconducting circuits},\ }\href {https://doi.org/10.1103/PhysRevLett.115.240502} {\bibfield  {journal} {\bibinfo  {journal} {Phys. Rev. Lett.}\ }\textbf {\bibinfo {volume} {115}},\ \bibinfo {pages} {240502} (\bibinfo {year} {2015})}\BibitemShut {NoStop}%
\bibitem [{\citenamefont {Zohar}\ \emph {et~al.}(2017)\citenamefont {Zohar}, \citenamefont {Farace}, \citenamefont {Reznik},\ and\ \citenamefont {Cirac}}]{zohar2017digital}%
  \BibitemOpen
  \bibfield  {author} {\bibinfo {author} {\bibfnamefont {E.}~\bibnamefont {Zohar}}, \bibinfo {author} {\bibfnamefont {A.}~\bibnamefont {Farace}}, \bibinfo {author} {\bibfnamefont {B.}~\bibnamefont {Reznik}},\ and\ \bibinfo {author} {\bibfnamefont {J.~I.}\ \bibnamefont {Cirac}},\ }\bibfield  {title} {\bibinfo {title} {Digital lattice gauge theories},\ }\href {https://doi.org/10.1103/PhysRevA.95.023604} {\bibfield  {journal} {\bibinfo  {journal} {Phys. Rev. A}\ }\textbf {\bibinfo {volume} {95}},\ \bibinfo {pages} {023604} (\bibinfo {year} {2017})}\BibitemShut {NoStop}%
\bibitem [{\citenamefont {Davoudi}\ \emph {et~al.}(2021)\citenamefont {Davoudi}, \citenamefont {Raychowdhury},\ and\ \citenamefont {Shaw}}]{davoudi2021search}%
  \BibitemOpen
  \bibfield  {author} {\bibinfo {author} {\bibfnamefont {Z.}~\bibnamefont {Davoudi}}, \bibinfo {author} {\bibfnamefont {I.}~\bibnamefont {Raychowdhury}},\ and\ \bibinfo {author} {\bibfnamefont {A.}~\bibnamefont {Shaw}},\ }\bibfield  {title} {\bibinfo {title} {Search for efficient formulations for hamiltonian simulation of non-abelian lattice gauge theories},\ }\href {https://doi.org/10.1103/PhysRevD.104.074505} {\bibfield  {journal} {\bibinfo  {journal} {Phys. Rev. D}\ }\textbf {\bibinfo {volume} {104}},\ \bibinfo {pages} {074505} (\bibinfo {year} {2021})}\BibitemShut {NoStop}%
\bibitem [{\citenamefont {Calaj{\'o}}\ \emph {et~al.}(2024)\citenamefont {Calaj{\'o}}, \citenamefont {Magnifico}, \citenamefont {Edmunds}, \citenamefont {Ringbauer}, \citenamefont {Montangero},\ and\ \citenamefont {Silvi}}]{calajo2024digital}%
  \BibitemOpen
  \bibfield  {author} {\bibinfo {author} {\bibfnamefont {G.}~\bibnamefont {Calaj{\'o}}}, \bibinfo {author} {\bibfnamefont {G.}~\bibnamefont {Magnifico}}, \bibinfo {author} {\bibfnamefont {C.}~\bibnamefont {Edmunds}}, \bibinfo {author} {\bibfnamefont {M.}~\bibnamefont {Ringbauer}}, \bibinfo {author} {\bibfnamefont {S.}~\bibnamefont {Montangero}},\ and\ \bibinfo {author} {\bibfnamefont {P.}~\bibnamefont {Silvi}},\ }\bibfield  {title} {\bibinfo {title} {Digital quantum simulation of a (1+ 1)d su(2) lattice gauge theory with ion qudits},\ }\href {https://doi.org/10.1103/PRXQuantum.5.040309} {\bibfield  {journal} {\bibinfo  {journal} {PRX Quantum}\ }\textbf {\bibinfo {volume} {5}},\ \bibinfo {pages} {040309} (\bibinfo {year} {2024})}\BibitemShut {NoStop}%
\bibitem [{\citenamefont {Robin}\ and\ \citenamefont {Savage}(2024)}]{robin2024magic}%
  \BibitemOpen
  \bibfield  {author} {\bibinfo {author} {\bibfnamefont {C.~E.}\ \bibnamefont {Robin}}\ and\ \bibinfo {author} {\bibfnamefont {M.~J.}\ \bibnamefont {Savage}},\ }\href@noop {} {\bibinfo {title} {The magic in nuclear and hypernuclear forces}} (\bibinfo {year} {2024}),\ \Eprint {https://arxiv.org/abs/2405.10268} {arXiv:2405.10268 [nucl-th]} \BibitemShut {NoStop}%
\bibitem [{\citenamefont {White}\ and\ \citenamefont {White}(2024)}]{white2024magic}%
  \BibitemOpen
  \bibfield  {author} {\bibinfo {author} {\bibfnamefont {C.~D.}\ \bibnamefont {White}}\ and\ \bibinfo {author} {\bibfnamefont {M.~J.}\ \bibnamefont {White}},\ }\href@noop {} {\bibinfo {title} {The magic of entangled top quarks}} (\bibinfo {year} {2024}),\ \Eprint {https://arxiv.org/abs/arXiv:2406.07321} {arXiv:arXiv:2406.07321 [hep-ph]} \BibitemShut {NoStop}%
\bibitem [{\citenamefont {Smith}\ \emph {et~al.}(2024)\citenamefont {Smith}, \citenamefont {Papić},\ and\ \citenamefont {Hallam}}]{smith2024nonstabilizerness}%
  \BibitemOpen
  \bibfield  {author} {\bibinfo {author} {\bibfnamefont {R.}~\bibnamefont {Smith}}, \bibinfo {author} {\bibfnamefont {Z.}~\bibnamefont {Papić}},\ and\ \bibinfo {author} {\bibfnamefont {A.}~\bibnamefont {Hallam}},\ }\href@noop {} {\bibinfo {title} {Non-stabilizerness in kinetically-constrained rydberg atom arrays}} (\bibinfo {year} {2024}),\ \Eprint {https://arxiv.org/abs/2406.14348} {arXiv:2406.14348 [quant-ph]} \BibitemShut {NoStop}%
\end{thebibliography}
\end{document}